\documentclass[12pt,preprint]{aastex}

\usepackage{lscape}

\newcommand{\cgs}{ erg cm$^{-2}$ s$^{-1}$ }

\newcommand{\sax}{{\it BeppoSAX} }
\newcommand{\asca}{{\it ASCA}}
\newcommand{\xmm}{{\it XMM-Newton} }
\newcommand{\xte}{{\it RossiXTE} }
\newcommand{\chandra}{{\it Chandra} }
\newcommand{\rl}{radio-loud~}
\newcommand{\rqu}{radio-quiet~}
\newcommand{\tnm}{\tablenotemark}
\newcommand{\nd}{\nodata}
\newcommand{\nh}{$N_{\rm H}\;$}
\newcommand{\mcl}{\multicolumn}

\shorttitle{BeppoSAX View of Radio-Loud AGN}
\shortauthors{P. Grandi, G. Malaguti, \& M. Fiocchi}

\begin{document}


\title{BeppoSAX view of radio-loud Active Galactic Nuclei}


\author{Paola Grandi, Giuseppe Malaguti  }
\affil{Istituto di Astrofisica Spaziale e Fisica Cosmica IASF/INAF - Bologna, Via Gobetti 101, Bologna, I-40129 Italy}
\author{Mariateresa  Fiocchi}
\affil{Istituto di Astrofisica Spaziale e Fisica Cosmica IASF/INAF - Roma , Via Fosso del Cavaliere 100, Roma, I-00133, Italy} 
\affil{Universit\`a La Sapienza, Piazza A. Moro, 5 I-00100 Roma, Italy}
\email{grandi@bo.iasf.cnr.it, malaguti@bo.iasf.cnr.it,fiocchi@rm.iasf.cnr.it}


\begin{abstract}
A systematic analysis of a large sample of radio-loud Active Galactic Nuclei (AGN) available in the {\em BeppoSAX} public archive 
has been performed. The sample includes 3 Narrow Line Radio Galaxies (NLRG), 10 Broad Line Radio Galaxies (BLRG), 6 Steep Spectrum Radio Quasars (SSRQ), and 16 Flat Spectrum Radio Quasars (FSRQ). According to the unified models, these classes correspond to objects with increasing 
viewing angles. As expected, the presence of a non-thermal beamed component emerges clearly in FSRQ.
This class shows in fact a featureless continuum (with the exception of 3C273), and a significantly flatter average spectral slope.
On the contrary, traces of a non-thermal Doppler enhanced radiation are elusive in the other classes. 
We find that the iron line equivalent widths (EW) are generally weaker in radio-loud AGN than in Seyfert 1 galaxies, and confirm  
the presence of an X-ray Baldwin effect, i.e. a decrease of EW with the 2--10 keV luminosity from Seyferts to BLRG and quasars. 
Since the EW--L$_{\rm 2-10\;keV}$ anti-correlation is present also in radio-quiet AGN alone, this effect cannot be ascribed entirely to 
a strongly beamed jet component. Possible alternative interpretations are explored. 
\end{abstract}


\section{Introduction}

The Active Galactic Nuclei (AGN) \rl/\rqu separation is essentially based on the radio property of the PG sample quasars \citep{kel89}. 
The ratio, $RL$, of radio-to-optical luminosity ($RL= \log [F_{\rm 5\;GHz}/F_{\rm B}]$) indicates the presence of two separates classes, 
with only $15-20$\% of the quasars much brighter ($RL>10$) at radio than at optical wavelengths. 
Recent deeper radio surveys associated with the high resolution radio continnum observations of radio quiet AGN are now changing this view. 
The FIRST survey has shown a smooth $RL$ distribution, instead of a bimodality \citep{whi00}. 
Moreover, non-thermal radio emission has been discovered also in Seyferts on parsec scales, although characterized by radio components generally 
non relativistic and weak \citep[e.g.:][]{ulve84a,ulve84b,ulve89,kuk95,na99,th01,mid04,ulve05}. 
A continuous distribution of radio--loudness rather than a sharp division between \rl and \rqu AGN seems to be the more convincing interpretation of the new observations. However, the wide range of the observed jet kinetic power still suggests the existence of differences among the various AGN classes and stimulates the search for physical reasons which can explain their variety.
Several possible solutions have been proposed so far. A fundamental parameter, able to control the AGN radio power, could be the black hole mass \citep{fra98}, even if there is not a general agreement on this hypothesis. \citet{mcl04} have recently confirmed the presence of a strong correlation between the AGN black hole mass and the 
radio-loudness $RL$ parameter, in disagreement with the result by \citet{woo02}, who, on the basis of a large compilation of black hole masses, arrived to opposite conclusions.
Another possible parameter responsible for the \rl and \rqu ``dichotomy'' could be the black hole spin, as originally suggested (``spin paradigm'') by \citet{bla90}. Jet production mechanisms, which involve the electrodynamic extraction of rotational energy from a Kerr black hole, could explain why some sources are \rl while some other are \rqu \citep{mei99,mei03}. However, also the spin paradigm has been questioned by recent works. For example, \citet{vol05}, having computed the expected distribution of massive black hole spins versus cosmic time, deduced that most black holes rotate rapidly at all epochs. This implies that the spin cannot be the only relevant parameter determining AGN radio loudness. As a consequence, the authors suggest that it is the mode of accretion, perhaps in addition to the spin, to determine the radio loudness. Actually, it is not yet clear if \rl and \rqu share the same accretion flow properties.  

In this respect, X-ray observations, which better probe the nature of the accretion, can play a fundamental role. \sax and \xte have shown that BLRGs have weak and narrow iron lines \citep{era00,gra01,zdz01} when compared to Seyfert 1 galaxies.  This could be an effect due to \rl/\rqu different disk geometries and/or accretion flow efficiencies (at least in the inner disk regions), or simply reflect the jet contamination of the Seyfert-like continuum.

Recent \xmm observations of two high luminosity \rl AGN still do not firmly clarify the issue. In 3C 111 \citep{lew05}, the spectrum presents broad residuals around the Fe K$\alpha$ line, but the precise properties of the line are highly dependent upon the chosen continuum model.  Evidence of a broad line is instead reported for 4C +74.26, which the authors propose to be consistent with the emission from the inner disk very close to a spinning black hole \citep{bal05}. On the other hand, in 3C 120 data indicate the presence of a narrow Fe line produced far from the black hole in the outer disk or in the broad line region \citep{ogl05,bal04}. 

Finally, a recent work on the FSRQ 3C 273 has shown that the jet and accretion flow component can have comparable intensity in the 2--10 keV range, and explained the observed weakness of the Fe K$\alpha$ line in terms of dilution of the disk emission caused by the jet component \citep{gra04}. This finding renews the necessity to understand the importance of the jet in the observed X-ray spectrum of \rl AGN.

In the present work an X-ray study of \rl AGN observed with the \sax satellite is presented.  The key aim is to assess the importance of jet emission exploiting the broad band high energy coverage ($\sim 0.1-200$ keV) capability of \sax. In particular we investigate possible jet contamination effects on the resulting X-ray spectrum for the various radio loud AGN classes, which correspond, within the unified schemes, to different viewing angles. In the effort to clarify the radio-loud/radio-quiet problem, the overall statistical properties of the present sample are compared with the data obtained for Seyfert galaxies by the same satellite \citep{per02}.

\section{The sample}

In the unified model scenario \citep{urr95}, type 1 (bright optical/UV continua and broad emission lines) Radio Loud AGN are 
classified as Broad Line Radio Galaxies (BLRG), or as Steep Spectrum Radio Quasars (SSRQ) at low and high luminosities, respectively. In particular, BLRG are usually considered the \rl counterparts of Seyfert 1 galaxies. Flat Spectrum Radio Quasars (FSRQ) are normally associated with very low inclination angle quasars, and show blazar properties. Narrow Line Radio Galaxies (NLRG) are type 2 AGN (weak optical/UV continua and narrow emission lines), and correspond to \rqu Seyfert 2 galaxies.

In order to sample all the possible inclination angles, we have considered all FSRQ, SSRQ, BLRG and  NLRG for which spectroscopical information is present in at least one of \sax Narrow Field Instruments (NFI): LECS \citep{par97}, MECS \citep{boe97}, and PDS \citep{fro97}.  The resulting sample includes 35 AGN (10 BLRG, 3 NLRG, 6 SSRQ, 16 FSRQ) for a total of 61 observations.  Most of the radiogalaxies (9) show a Fanaroff-Riley II (FR II) morphology, one source, 3C120, is a FR I, while three objects have uncertain/unknown radio morphology.

\begin{figure}
\epsscale{.80}
\plottwo{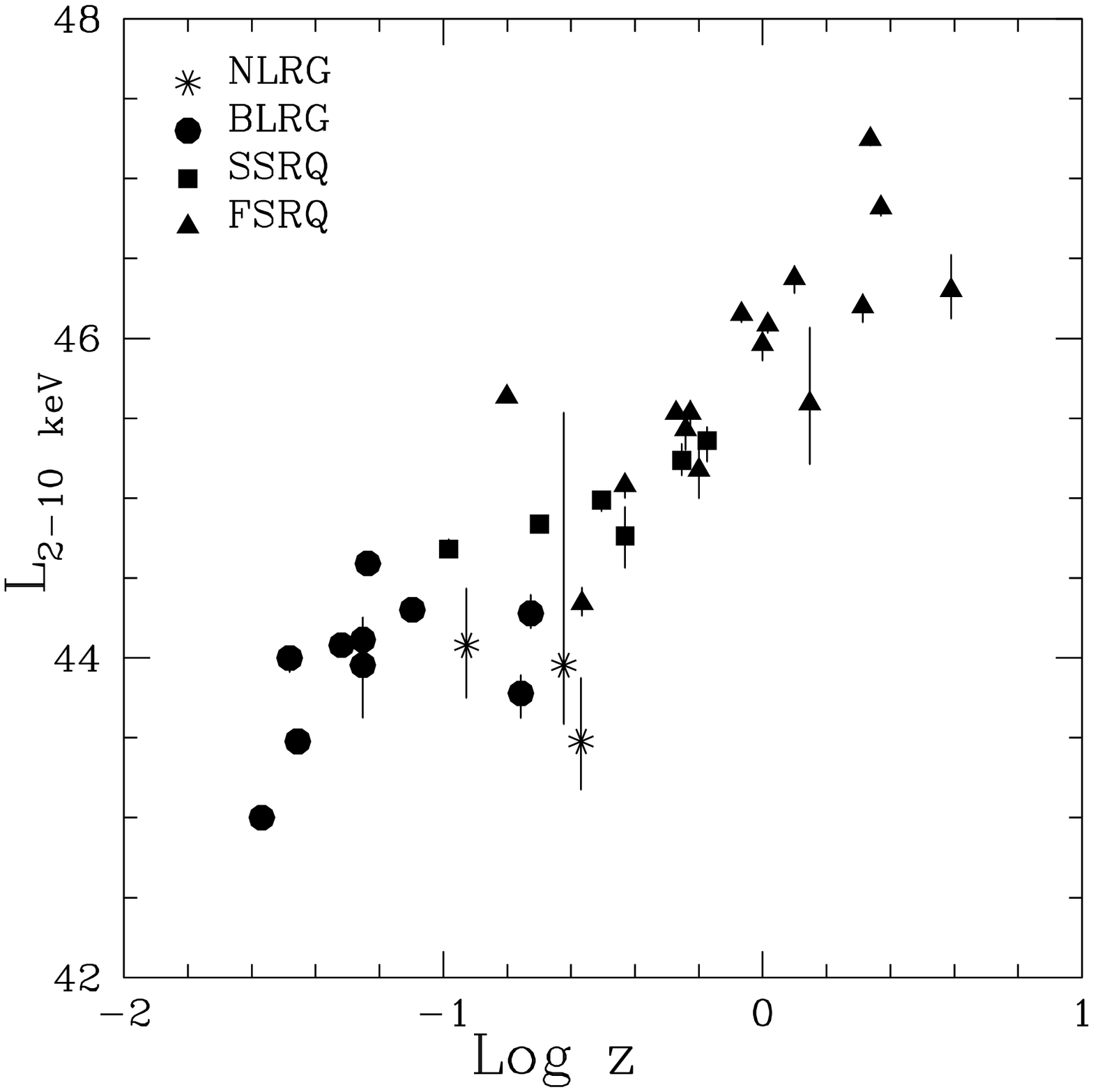}{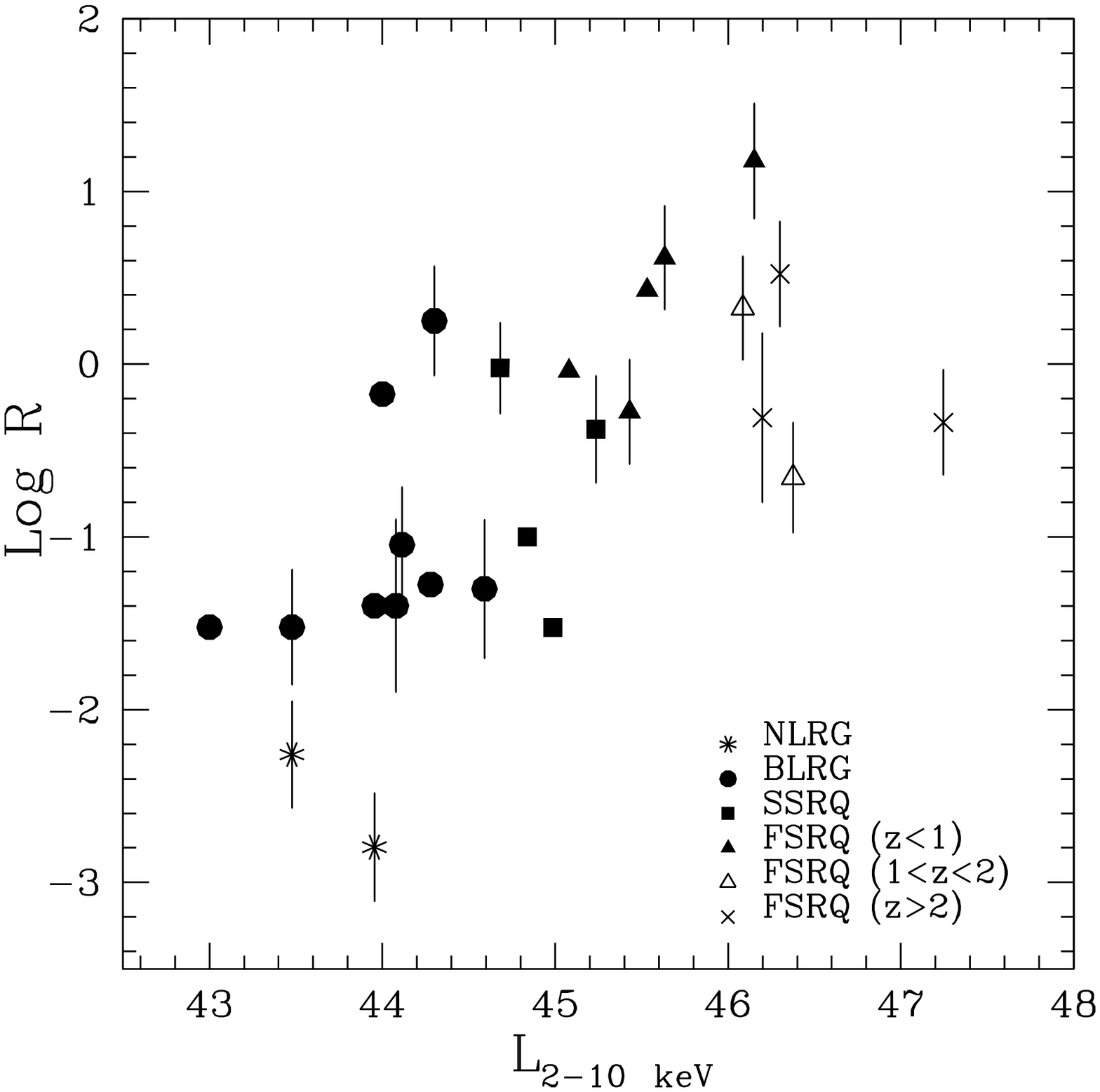}
\caption{{\it Left Panel} -- X-ray Luminosity in the 2-10 keV band (rest frame) versus redshift.
{\it Right Panel} -- X-ray Luminosity in the 2-10 keV band (rest frame) versus Radio Core Dominance. In agreement with the Unified Schemes, R increases
from NLRG to FSRQ}.
\label{fig:Lum}
\end{figure}

The sample, shown in Figure \ref{fig:Lum} (left panel), spans about four orders of magnitude in X-ray luminosity ($L_{\rm 2-10\;keV}\sim10^{43}\div10^{47}$ erg $s^{-1}$), and about two in redshift ($z=0.027\div3.9$). Table \ref{tab:sam} lists the relevant information of all the analyzed AGN. The Radio Core Dominance value, $R$, reported in Table \ref{tab:sam} is defined as $R=S_{\rm core}/[S_{\rm tot}-S_{\rm core}]$, being $S_{\rm core}$ and $S_{\rm tot}$ the core and total flux densities at 5 GHz, respectively. 
$R$ is considered an indicator of the orientation of the beamed radiation with respect to the line of sight \citep{orr82}. In agreement with the unified schemes, objects with large $R$ are expected to have their beamed radiation emitted in a direction closer to the line of sight, and therefore with a higher beam boosting, even if the X-rays from the jet are not necessarily beamed as the radio ones. 
A study of radio cores of a complete subsample of radio sources taken from the 2.7 GHz \citet{wap85} sample \citep{mor97} has indeed 
shown that $R$ increases going from NLRG to FSRQ, with BLRG between NLRG and SSRQ.
 
The values of $R$ in Table \ref{tab:sam} are from \citet{mor93} and \citet{fan03}.
\citet{fan03} calculated $R$ at 5 GHz using extended luminosities at 1.4 GHz.
They  assumed two possible values of the extended spectral index ($\alpha=0.5$ and $\alpha=1.0$) to convert the 1.4 GHz luminosity into 5 GHz luminosity and gave two values of $R$ for each analyzed source. Table \ref{tab:sam} shows the average value $R=(R_{\alpha=0.5}+R_{\alpha=1})/2$, together with the associated $\sigma_{\rm R}=|R_{\alpha=0.5}-R_{\alpha=1}|/2$, which represents the allowed range covered by $R$. 

Figure \ref{fig:Lum} (right panel) shows, the different classes of the sample on the $L_{\rm 2-10\;keV}-R$ plane. 
In agreement with the unified schemes, there is a general increasing trend of $R$, i.e. of the contribution of the Doppler-enhanced non-thermal radiation, from Radio-Galaxies to Quasars. FSRQ occupy the upper region above $R>0.1$, the two NLRG are in the lower region of the plane, and most BLRG are in the intermediate region, while SSRQ are equally distributed in the FSRQ and BLRG regions.

\section{Data Reduction and Spectral Analysis}

Table \ref{tab:jou} summarizes the log of \sax observations.  LECS, MECS and PDS event files and spectra, available from the ASI Scientific Data Center (ASDC), were generated by means of the Supervised Standard Science Analysis \citep[~FGG hereafter]{fio99}. For the PDS, we have generally used the spectra obtained with the the background rejection method based on fixed rise time selection. When possible the spectra obtained with variable rise time selection techniques were used in order to improve the statistics (FGG). Spectra having low statistics 
(S/N ratio $<3$) or being background contaminated have not been included in the dataset, and are marked with ``N.S.'' in Table \ref{tab:jou}. The only exception is 3C 445, the PDS flux of which is contaminated by the bright Abell cluster A2440 ($z=0.09$) located only $30'$ apart.  Since the X-ray emission of this cluster is known, $F_{\rm 2-10\;keV} \sim 1.22\times10^{-11}$ erg cm$^{-2}$ s$^{-1}$, kT$\sim$9 keV \citep{dav93}, its contribution was taken into account in the 3C 445 fit. The flux of the cluster was extrapolated to the 15--200 keV range and rescaled for the off-axis PDS effective area at $30'$ from the field of view center.

Spectral data were then binned using the template files distributed by ASDC, in order to ensure the applicability of the $\chi^2$ statistic and an adequate sampling of the spectral resolution of each instrument. For each source, data from different instruments were simultaneously fitted with XSPEC v. 11.1. Different relative normalization constants were used to take into account mis-calibrations among the instruments. Inter-calibration constant values were taken in agreement with the indications given in FGG.

To verify fast flux variations (hours/days) within a single pointing, a light curve was extracted for each observation and a $\chi^2$ test applied against a constant hypothesis. Although fast variability ($\chi^2$ probability $\le10^{-3}$) was observed in several sources, limited statistics did not allow time-resolved spectral analysis. Therefore spectral fits were always performed on the entire observations. For the same reason, short ($<30$ ks) pointings of selected sources (3C 57, 0528+134, 3C 273, 3C279, PKS 2230+114) repeated every few days, were combined together.

Long (months, years) time flux variability was also checked.  Again, for the sources showing no significant luminosity/spectral changes, spectra from different observations were summed together to improve statistics. The only exception is 3C 273, for which a detailed variability study has been recently reported \citep{gra04}. In the present work, we discuss the average spectral properties of 3C 273, analyzing only the aggregate spectrum obtained combining all observations.

Each source was initially fit with a power-law absorbed by galactic column density. When statistically required, more complex models were used including a high energy cutoff, $E_{\rm C}$, and a reflection component ($Refl.$). Each time a new component was added to the model, a F-test was performed. We assumed that a F probability larger than 95\% implies a significative improvement of the fit.

The \textsc{pexrav} code \citep{mag95} in XSPEC was used to fit the direct continuum and the reflected component. The inclination angle ($i$) of the reflecting cold matter was always fixed to $i=18^\circ$ ($\cos\;i = 0.95$). Although a wider range of inclination angles is probably covered by the analyzed type 1 AGN, the face-on geometry was chosen to allow for a coherent comparison with Seyfert 1s results \citep{per02}. The spectrum of NLRG was only fit with a power law because of the lack of PDS signal.

The presence of a soft excess and of an additional (intrinsic) \nh was also systematically searched for. 
Since the limited LECS statistics did not allow to distinguish between the different possible physical origins of the observed soft excesses, a thermal model (\textsc{mekal} in XSPEC) was applied to all sources. In only one case, 3C 273, the soft excess was fit with a black body model to account for the possible accretion disk emission suggested by the observed strong UV bump \citep{ulr81}.

The presence of a fluorescence iron line was also investigated. 
We looked for an excess in emission with respect to the continuum between 5.5 and 7.5 keV (rest frame).
When the statistical significance of this excess, estimated by the quadratic sum of the deviations, was larger than $3\sigma$, a gaussian line
was added to the model. Then, a F-test was also applied as further check. At first all parameters of the gaussian line were let free to vary. 
Since the detected iron lines were always consistent with a cold and unresolved feature, the final fits were performed keeping the 
energy centroid and the intrinsic width fixed to $E_{\rm Fe}=6.4$ keV and $\sigma_{\rm Fe}$=0 keV, respectively.

Throughout the paper, all errors and upper-limits are quoted at 90\% confidence for two interesting parameters ($\Delta\chi^2=4.61$). For calculations of luminosity, $H_0=75$ km s$^{-1}$ Mpc$^{-1}$ and $q_0=0.5$ are assumed.

\section{Spectral Analysis Results}

Spectral fit results, grouped for class, are summarized in Table \ref{tab:fit}. The parameters of the soft component, when required, are listed in Table \ref{tab:soft}. The $\chi^2$ values in Table \ref{tab:fit} refer to the total fit including, when applicable, the extra soft emission.\\

\subsection{BLRG}
In general a simple power law plus Galactic absorption model does not give good fits to the data. Absorption in excess of the Galactic column density, iron lines, reflection humps, soft excesses and steepening of the high-energy spectrum are common spectral features, although not always simultaneously present in each source. Although it appears evident that short exposures ($<50$ ks) were generally inadequate to detect reprocessed features, the observed spectral variety can not be entirely ascribed to statistical limits. For example, the two BLRG, 3C 111 and 3C 390.3, appear clearly different in spite of their similar 2--10 keV flux and \sax exposure time. The 3C 111 continuum results flat and only upper limits are available for the reprocessed features. On the contrary 3C 390.3 is steep with the presence of a prominent Fe feature ($EW\sim120$ eV) and a remarkable reflection component.

A column density in excess to the Galactic one is found in half of the BLRG (excluding 3C 111). However, the absorber is generally thin (N$_{\rm H}<10^{22}$ cm$^{-2}$). The only exception is 3C 445, which is heavily obscured by a complex dual absorber.  For 3C 111, the extra absorber is probably local to our Galaxy, and associated to a dense molecular cloud \citep{ban91}.

A soft excess emission is measured for only two sources (3C 120 and 3C 382). Although a thermal parameterization is used to account for this, it is not possible to exclude a nuclear origin of the soft photons. Several elements support this possibility. The soft X--ray spectrum of 3C 382 is very complex and requires two components, one of which can be non-extended \citep{gra01}. Recent observations of 3C 120 performed with XMM--Newton suggest the possibility of a point-like nature for the soft emission \citep{ogl05,bal04}. Finally, strong UV bumps have been observed in both sources \citep{mar91,tad86}.\\

\subsection{NLRG}
As expected within the unified scheme, NLRG are heavily absorbed. In two cases, because of poor statistics, the spectral index was fixed to the best fit value in order to better constrain the $N_{\rm H}$. In both cases the intrinsic absorber results significantly larger ($N_{\rm H}\simeq10^{23}$ cm$^{-2}$) than what measured for BLRG. For 3C 300, only an upper limit is obtained for the absorbing column. However, its flat spectral slope could indicate the presence of a larger absorber. The de-absorbed luminosity is within the range covered by the BLRG.\\

\subsection{SSRQ}
Short available exposures and low X-ray fluxes ($F_{\rm 2-10\;keV} < 10^{-12}$ \cgs) generally prevent the detection of spectral features. In only one case, 4C 74.26, the statistics allows a deeper analysis. The measured spectral features, $E_{\rm C}$, reflection, and FeK$\alpha$ EW turn out to be similar to the BLRG, even though less constrained. The data give indication for the presence of a broad line (see note $f$ in Table \ref{tab:fit}), as recently also suggested by \citet{bal05}, using XMM-Newton results.\\

\subsection{FSRQ}
The 0.1--200 keV spectrum is generally well described by a simple power law. The data do not require the addition of an iron line. Even for the objects having PDS data, it is possible to give only upper or lower limits to the reflection component and $E_{\rm C}$, respectively. The only exception is 3C 273. The combination of high statistics and repeated observations allows the detection of significant deviations from a simple power law model. These spectral features indicate the presence of accretion, although generally overwhelmed by a strong non-thermal beamed emission \citep{gra04}.

Blazars are known to be dominated by non-thermal Doppler enhanced emission. The generally hard spectral slopes ($\Gamma \sim 1.5-1.6$) indicate that all the analyzed FRII/FSRQ are in the Compton-regime in the medium-hard X-ray bands, as expected for high luminosity Blazars \citep{fos98}. Actually, synchrotron plus Compton emission models have been applied with success to several FSRQ of our sample \citep{tav00,don01,tav02}. A possible counterexample is RGB J1629+4008, which shows a very soft ($\Gamma\sim2.6$) spectrum. \citet{pad02} explained the X-ray spectrum of this object as dominated by synchrotron radiation, suggesting the existence of FSRQ sources showing properties similar to high energy peaked BL Lacs (HBL) objects. On the other hand, the optical classification as a NLS1 on the basis of its $H_{\rm \beta}$ line width \citep{gru04}, could explain the steep soft X--ray spectrum as due to thermal emission possibly from an accretion disk. Because of its peculiarity, this source is not included in the following statistical studies. 

Three quasars, namely NRAO 140, 3C 454.3 and 4C 71.07, need the addition of an extra absorber, which however is probably not intrinsic to the sources. In NRAO 140 and 3C 454.3 the absorber could be due to a molecular cloud in our Galaxy \citep{ban91,ung87,sam97}. In 4C 71.07, the extra \nh could be associated with a foreground damped Ly$\alpha$ absorber located at $z=0.3130$ \citep{tur03}.

\section{Searching for beaming effects in radio-loud AGN}

FSRQ are the AGN class showing significantly harder spectral slopes, when compared to BLRG and SSRQ.
According to the Kolmogorov-Smirnov test, the probability that the analyzed classes (FSRQ vs BLRG and FSRQ vs SSRQ) are drawn from the same distribution is very small (P$_{KS}=5\times10^{-3}$ and P$_{KS}=10^{-3}$, respectively).
This is not a new result. As FSRQ have the larger core dominance value ($R$), a strong hard non-thermal contribution is expected to affect their X-ray continuum. 

What is instead interesting to note is that, taking into account the whole sample, the X-ray spectral slope does not correlate with $R$. A non-parametric statistical test (Kendall's tau test) results in a correlation coefficient of $r_{\rm s}=-0.34$, with a significance value $\alpha_{\rm K}=0.1$, above the 0.05 limit that we have assumed to accept two quantities as correlated.
The search for a correlation between the high energy cutoff and the beaming indicator also gives a negative result. In order to assess the significance of the correlation in the presence of lower limits, we applied the generalized Kendall's tau test in the ASURV package \citep{iso86} to the total sample. The resulting significance, $\alpha_{\rm K}^{\rm ASURV}=0.2$, is again significantly higher than the acceptance threshold.

\begin{figure}
\epsscale{.80}
\plotone{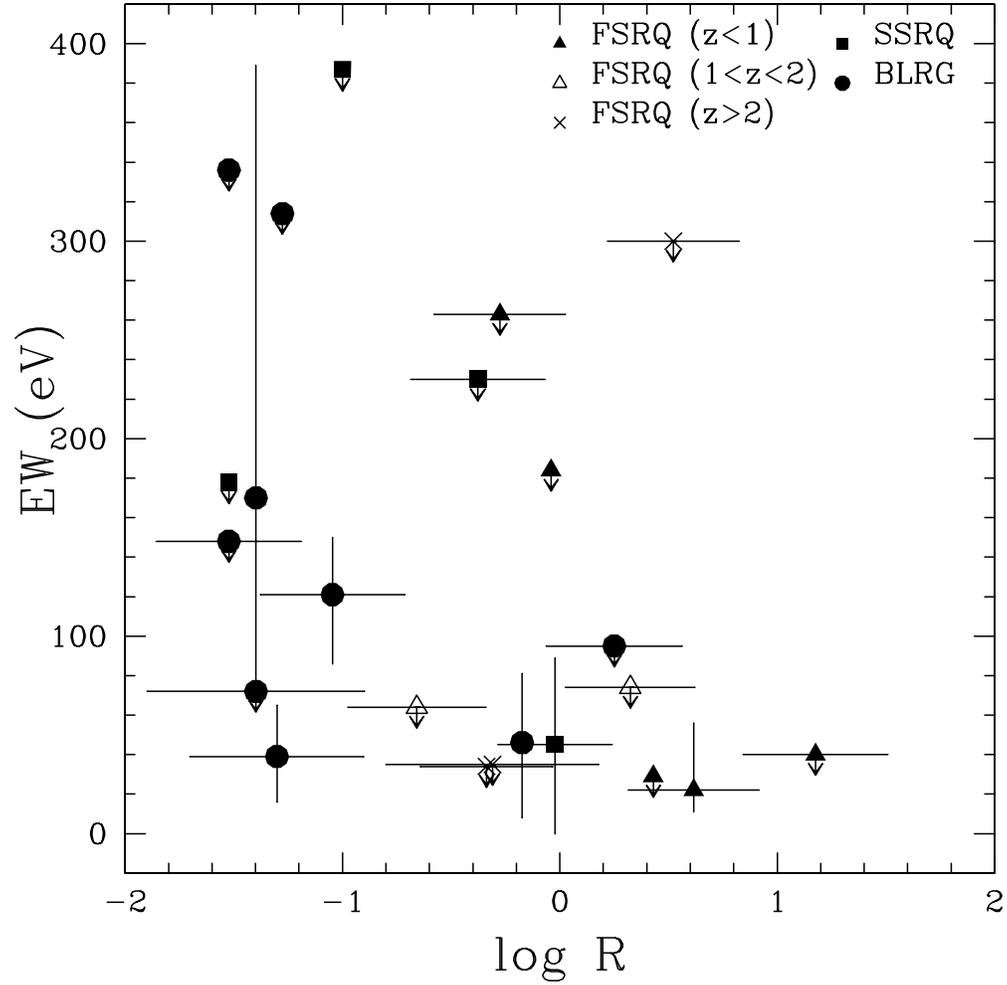}
\caption{Reprocessed components: iron line Equivalent Width plotted as a function of the Radio Core Dominance parameter for FSRQ, BLRG and SSRQ. A decreasing trend is apparent (see text).}
\label{fig:EWvsR}
\end{figure}

For the reprocessing features, iron line and reflection component, the situation is slightly different. There is an apparent decreasing trend of the equivalent width and the reflection component against the radio core dominance. The Kendall's tau test returned significance values of $\alpha_{\rm K}^{\rm ASURV}=0.07$ and $\alpha_{\rm K}^{\rm ASURV}=0.06$ for the EW vs $R$ and Ref vs $R$ correlations, respectively, in both cases only slightly above the 0.05 acceptance threshold. Figure \ref{fig:EWvsR} shows the EW against $R$ for FSRQ, BLRG and SSRQ. However, if we exclude the highly absorbed BLRG 3C 445 ($N_{\rm H}>10^{23}$ cm$^{-2}$), the already weak correlations disappear ($\alpha_{\rm K}^{\rm ASURV}>0.2$), casting farther doubts on the link between iron line EW and beaming. The intense iron line (EW$\sim170$ eV) observed in 3C 445, instead of being the result of a low beaming contamination, could be due to the contribution of the transmission through a line-of-sight absorber, as discussed above.

\begin{figure}
\epsscale{.80}
\plotone{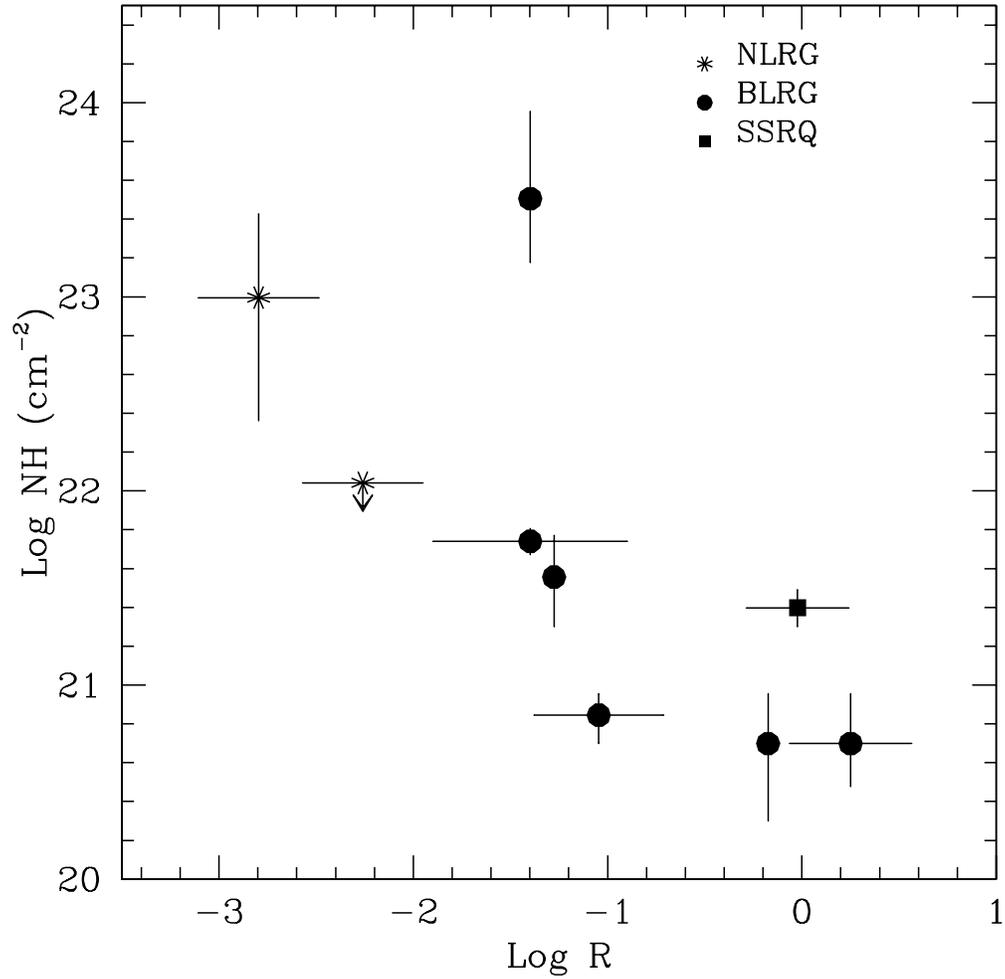}
\caption{Intrinsec column density (\nh) as a function of the Radio Core Dominance (R). In agreement with the unified models, the NLRG
continuum is more obscured and less beamed. 
As the ambiguous nature of the 3C 111 absorber, this source is not in the plot}.
\label{fig:NHvsR}
\end{figure}

The intrinsic column density \nh is the only spectral parameter correlating with the radio core dominance. The amount of gas along the line of sight decreases with increasing $R$ (Figure \ref{fig:NHvsR}). The generalized Kendall's tau test gives $\alpha_{\rm K}^{\rm ASURV}=0.02$. If we exclude all upper limit values, the significance further improves to $\alpha_{\rm K}=0.01$. This result is coherent with the unified model scenarios, which predict the nuclear source to be more obscured and the jet emission strongly de-amplified at large viewing angles.
Note that this correlation indirectly confirms the robustness of $R$ as indicator of the beaming orientation. 

The conclusion is that the lack of a strong correlation between $R$ and the X--ray spectral parameters reflects an intrinsic weakness 
of the jet contribution in most of the objects of our sample. This does not necessarily mean that the jet is not present. Our statistics could just be not sufficient to reveal its presence.  Only for the Blazars, the non-thermal emission is strong enough to clearly emerge in the observed X--ray spectrum (hard power law, absence/weakness of reprocessing features, no, or very high, $E_{\rm C}$).

\section{Radio-quiet versus radio-loud AGN}

\begin{figure}
\epsscale{.80}
\plottwo{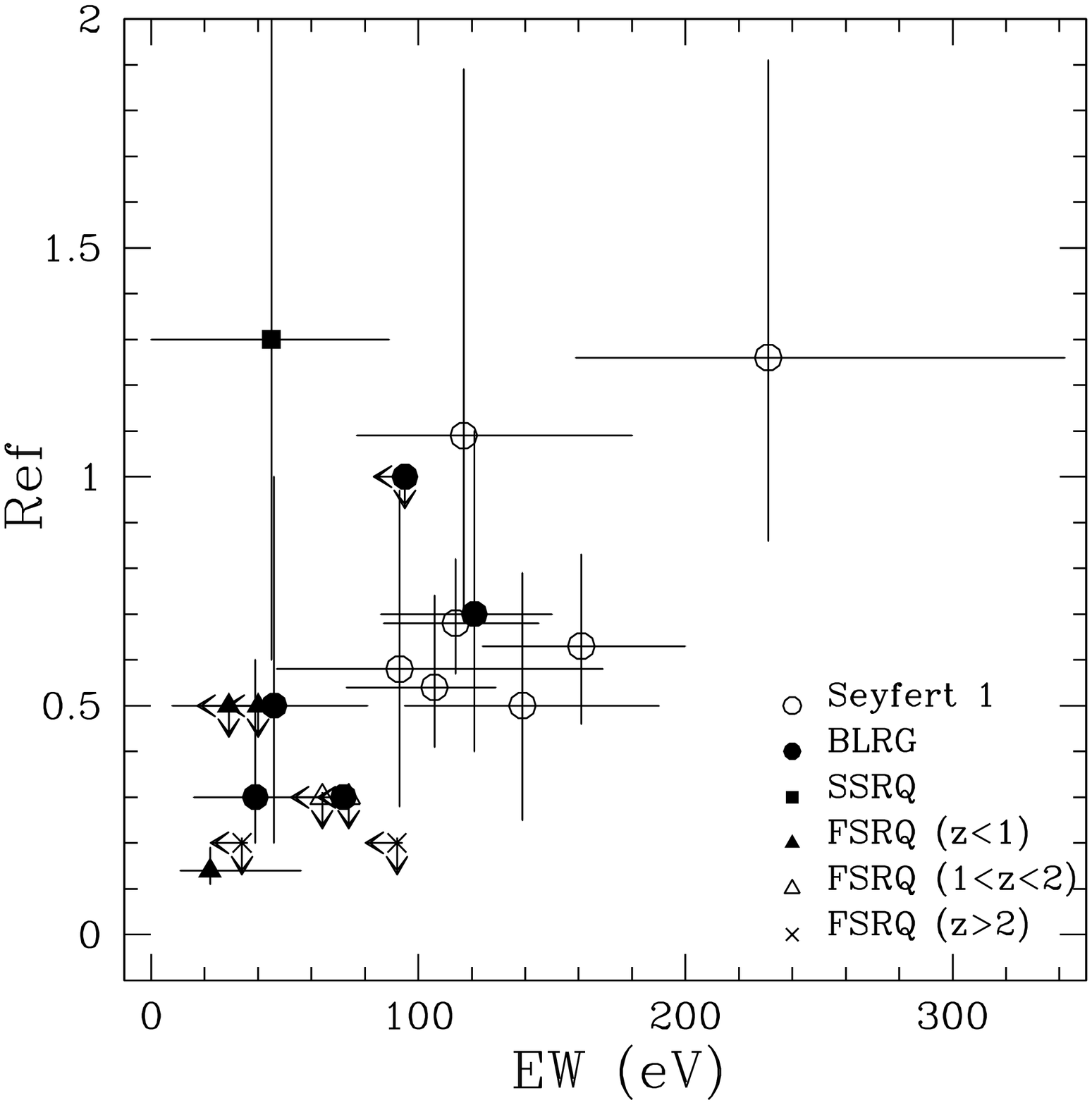}{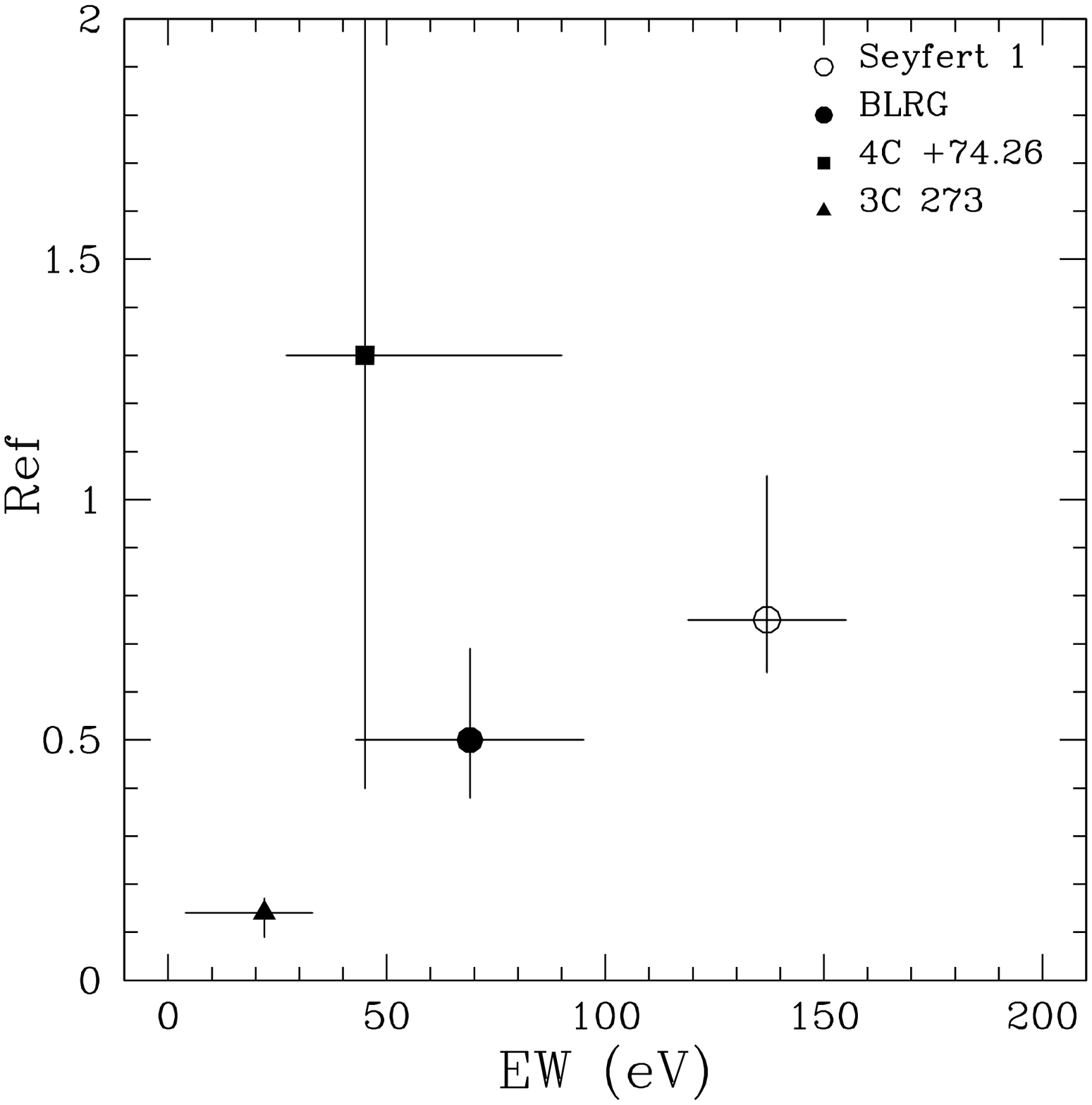}
\caption{{\it Left Panel} --Amount of reflected radiation (Ref)
plotted as a function of iron line strength (EW) for \rl AGN of our sample, excluding the strongly absobed source 3C445 (see text) .
For comparison the Seyfert 1 sample of Perola et al (2002) is also shown.
({\it Right panel}) -- Same plot with BLRG and Seyfert 1 average parameters (in Table 5).
Radio-loud AGN are generally characterized by weaker iron lines and reflection components}.
\label{fig:RefvsEW}
\end{figure}

Average X-ray properties for each class of the \rl sample are reported in Table \ref{tab:avg}. In order to compare our sample with \rqu objects, we included also the average spectral properties of the 7 Seyfert 1 galaxies analyzed by \citet{per02}. For this comparison we have chosen not to include 3C 445, because of its strong line-of-sight intrinsic absorption, in order to avoid objects in which part of the Fe line is produced via transmission \citep{ghi94}. The average continuum parameters ($\Gamma$, $E_{\rm C}$) indicate a strong similarity between BLRG and Seyfert 1, in contrast with Blazars. On the contrary, there is indication that the importance of reprocessed features smoothly increases from FSRQ to BLRG and Seyferts (we will return to
this point in Section 8).
Figure \ref{fig:RefvsEW} (left panel) shows the reflection component plotted against the Fe line EW for the whole sample plus Seyfert 1.  BLRG and FSRQ tend to occupy the lower left corner of the plot, and, taking into account the average values (Table \ref{tab:avg}), this effect becomes more evident (figure \ref{fig:RefvsEW}, right panel).

\section{Mass Accretion Rates}
\label{sec:mass}

Black hole masses ($M_{\rm BH}$) and bolometric luminosities ($L_{\rm Bol}$) were collected from the literature for all the sources in our sample and for Seyfert 1 galaxies of the \citet{per02} sample. For each source, the observed mass accretion rate in Eddington units has been computed as  $\dot{m}_{\rm obs}=L_{\rm Bol}/\epsilon L_{\rm Edd}$, having assumed a radiative efficiency $\epsilon=1$ in agreement with \citet{mar04}.  
The data are shown in Table \ref{tab:bhm}.
For each class (including Seyfert 1s) the average values, the standard errors on the mean, and the standard deviations, 
for $M_{\rm BH}$, $L_{\rm Bol}$ and $\dot{m}$ are also estimated and listed in Table \ref{tab:bhm}.
Our aim has been to investigate the nuclear properties of the objects of our sample looking for possible links with the 
observed X-ray spectral parameters.

A Kolmogorov-Smirnov test, applied to the \rl classes, does not show significant differences in the distribution of $M_{\rm BH}$. For each set of data, the associated probability that two arrays are drawn from the same distribution is $P_{\rm KS}>0.3$. 
The only exception are NLRG which show smaller $M_{\rm BH}$ when compared to quasars. 
However, given the small number of NLRG (only 3 objects), this result can not be considered very meaningful.

For the accretion rates, the results are different. We found that the $L_{\rm Bol}/L_{\rm Edd}$ ratio for FSRQ is always significantly larger than for BLRG ($P_{\rm KS}=3\times10^{-3}$) and SSRQ ($P_{\rm KS}=0.01$). Moreover a strong correlation ($r=0.72$) between $\dot{m}$ and $R$ is present ($\alpha_{\rm K}<0.01$) as appears evident in Figure \ref{EffvsR} (left panel). However we individuate two possible effects inducing the (not necessary real) correlations.
(1) The Doppler enhanced, non-thermal, emission, expected to increase with $R$, could shift the points (in particular FSRQ) to the upper region of the plot, because of an overestimation of $L_{\rm Bol}$. 
In addition, (2) the accretion rate value of NLRG could be affected by gas obscuring the nucleus. 
The bolometric luminosity in Table \ref{tab:bhm} has been derived from the optical luminosity using a conversion factor inferred by the Quasar Spectral Energy Distribution study by Elvis et al. (1994). The ratio between $<L_{\rm Bol}>$ and $<L_{\rm 2-10\;keV}>$ indicates that the X-ray conversion factor is very similar to the BLRG optical one, but much smaller in the case of NLRG. This could reflect uncertainties in the optical de-reddening.
Indeed, if the X-ray conversion factor is applied to the average NLRG X-ray luminosity value, the deduced $<L_{\rm Bol}>$ increases by more than a factor 10, and, as a consequence, the accretion rate  of NLRG aligns with the BLRG value, increasing from 
$\log{\dot{m}}=-2.92$ to $\log{\dot{m}}=-1.4$.

On the contrary, the statistical difference between Seyfert 1s and BLRG accretion rates appears to be more genuine. Seyfert 1s seem to have more 
efficient accretion, as confirmed by the Kolmogorov-Smirnov test ($P_{\rm KS}=3\times10^{-3}$). Note that, if the $L_{\rm Bol}$ of radio galaxies is affected by beamed radiation, the discrepancy becomes larger.

\section{X-ray Baldwin effect}

\label{sec:bal}
\begin{figure}
\epsscale{.80}
\plotone{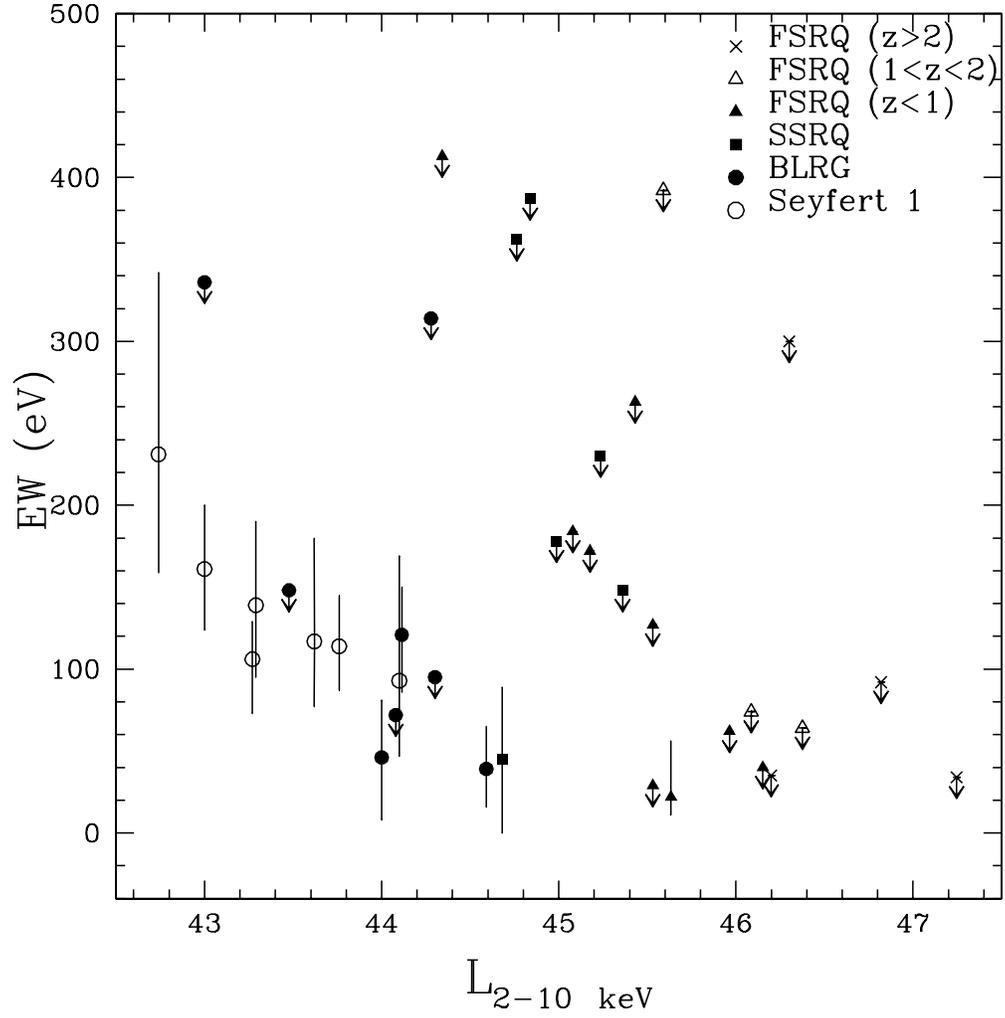}
\caption{Iron line EW versus X-ray luminosity between 2-10 keV.
An anti-correlation is statistically confirmed by the generalized Kendall's tau test}.
\label{EWvsL}
\end{figure}

Figure \ref{EWvsL} shows a clear anti-correlation between the iron line EW and the 2--10 keV luminosity, the well known ``X-ray Baldwin effect''. As confirmed by the generalized Kendall's tau test, a correlation is present, not only when both \rqu and \rl objects are 
considered ($\alpha_{\rm K}^{\rm ASURV}<0.01$), but also when Seyfert 1s are removed from the sample ($\alpha_{\rm K}^{\rm ASURV}=0.04$). The simplest and immediate interpretation is that a beamed radiation dilutes the features in \rl AGN. However, while for Blazars the robustness of this interpretation is clearly attested by the case of 3C 273 \citep{gra04}, for the other \rl AGN the question is not so certain (as shown in Section 5). 
Moreover, the observation of the Baldwin effect also in radio quiet AGN \citep{pag04,zho05} indicates a possibly more complex picture.
We then investigated other possibilities.

Most of the following speculations are based on the results reported in section \ref{sec:mass} where the accretion rate of BLRG and Seyferts are compared. We are aware that our sample is small and far from to be complete. However, we note that the average accretion rate obtained in this work for BLRG is consistent with what obtained by \citet{mar04}, using a larger sample. We also computed, as a further check, the average accretion rate of 17 Seyfert 1 galaxies reported in the AGN compilation by \citet{woo02}, for which it was possible to obtain the masses via reverberation mapping method. Again we find a substantial agreement ($Log~ \dot{m}=-0.79\pm0.18$) with the average Seyfert 1 value reported in Table \ref{tab:bhm}.

\subsection{Hot accretion flow}

An idea long debated in the past is that BLRG contain a hot accretion flow in contrast to the cold optically thin disk proposed for Seyferts \citep{haa91,haa93}. 
\citet{ree82} speculated on the possibility that in radio-galaxies the hot accreting gas is in the shape of an ion-supported torus characterized by low radiative efficiency. The advection dominated accretion flow (ADAF) models \citep{nar98}, successively proposed, follow similar lines of thought. In this picture, the accretion flow in BLRG could be hot and geometrically thick in the inner regions and become cold and geometrically thin only at larger radii \citep{che89,gra99,era00}. The weakness of the reprocessed features would be immediately explained by the small solid angle subtended by the cold matter (the external disk) to the primary X-ray source (the ion--supported torus). In that case a jet dilution of the X-ray continuum is not required anymore. 

Our analysis indicates the existence of smaller accretion rates in BLRG when compared to Seyfert 1s, supporting the idea that \rqu AGN may host accretion flow mechanisms different from \rl AGN. However, after the launch of XMM-Newton and \chandra, our view of \rqu AGN became more confused and this interpretation less appealing. The broad Fe lines seem not to be so common or strong as it was indicated by the early ASCA observations \citep{nan97a} and, indeed, a  Fe line red tail was not observed in most of the Seyferts of the Perola sample: IC 4329a \citep{ste05}, NGC 5548 \citep{pou03}, NGC 7469 \citep{blu03}, NGC 4593 \citep{rey04}.

\subsection{Cold accretion disk}

Small reprocessed features do not necessarily imply the disruption of the cold thin disk in its inner regions, but could indicate, for example, the presence of a highly ionized disk \citep{ros99}. \citet{nay00} investigated how the ionization of increasingly deep layers of a thin optically thick disk can affect the reprocessed features. In particular they showed that if the X-rays are produced in magnetic
flares, then the Fe line EW is indeed expected to decrease with increasing accretion rate. 
Although attractive, this hypothesis is however in clear disagreement with our results. BLRG, which show weaker lines, are characterized by less efficient accretion rates with respect to Seyfert 1 galaxies.

An alternative viable explanation, which preserves the presence of cold disks in \rqu and \rl AGN, has been recently suggested by \citet{min04}. On the basis of a light bending model, they demonstrate that sources with similar intrinsic luminosities can show different X-ray fluxes depending on the height of the primary X-ray source above the accretion disk. In particular, if the X-ray continuum is produced at low heights, 
the observed flux is lower and, taking into account relativistic effects, significatively beamed along the equatorial plane.
Larger EWs are then expected when the source is in a low state. The X-ray photons, preferably emitted along the 
equatorial pane, can hit outer regions of the disk, the torus and/or the broad line region, producing an enhancement of the 
(narrow) Fe feature. 
In this view, the Baldwin effect observed in Seyfert 1s and BLRG could be easily interpreted in terms of a larger height of the primary X-ray source in BLRG, without invoking the presence of a strong jet emission. 

\subsection{Molecular torus}

\begin{figure}
\epsscale{.80}
\plottwo{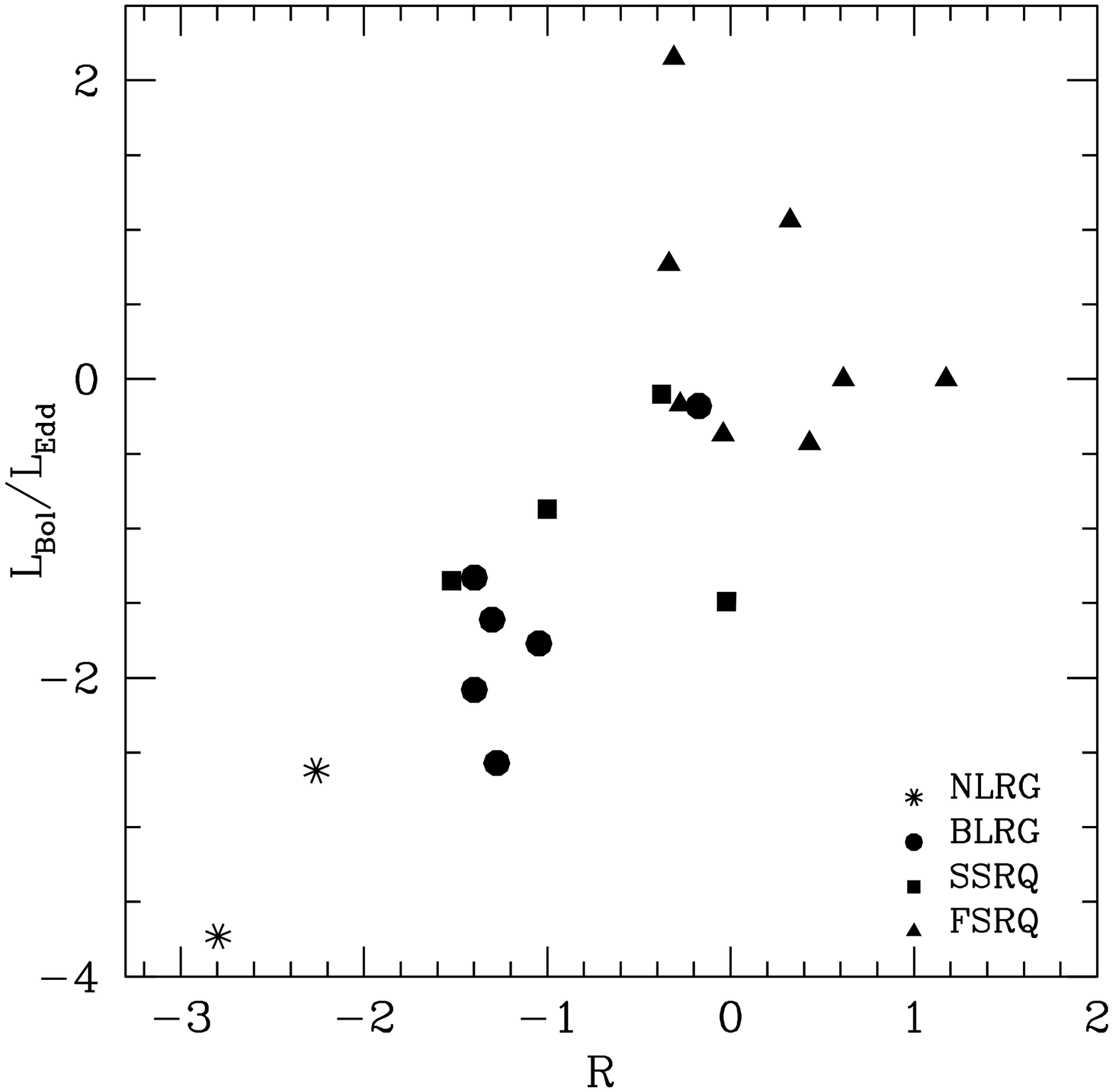}{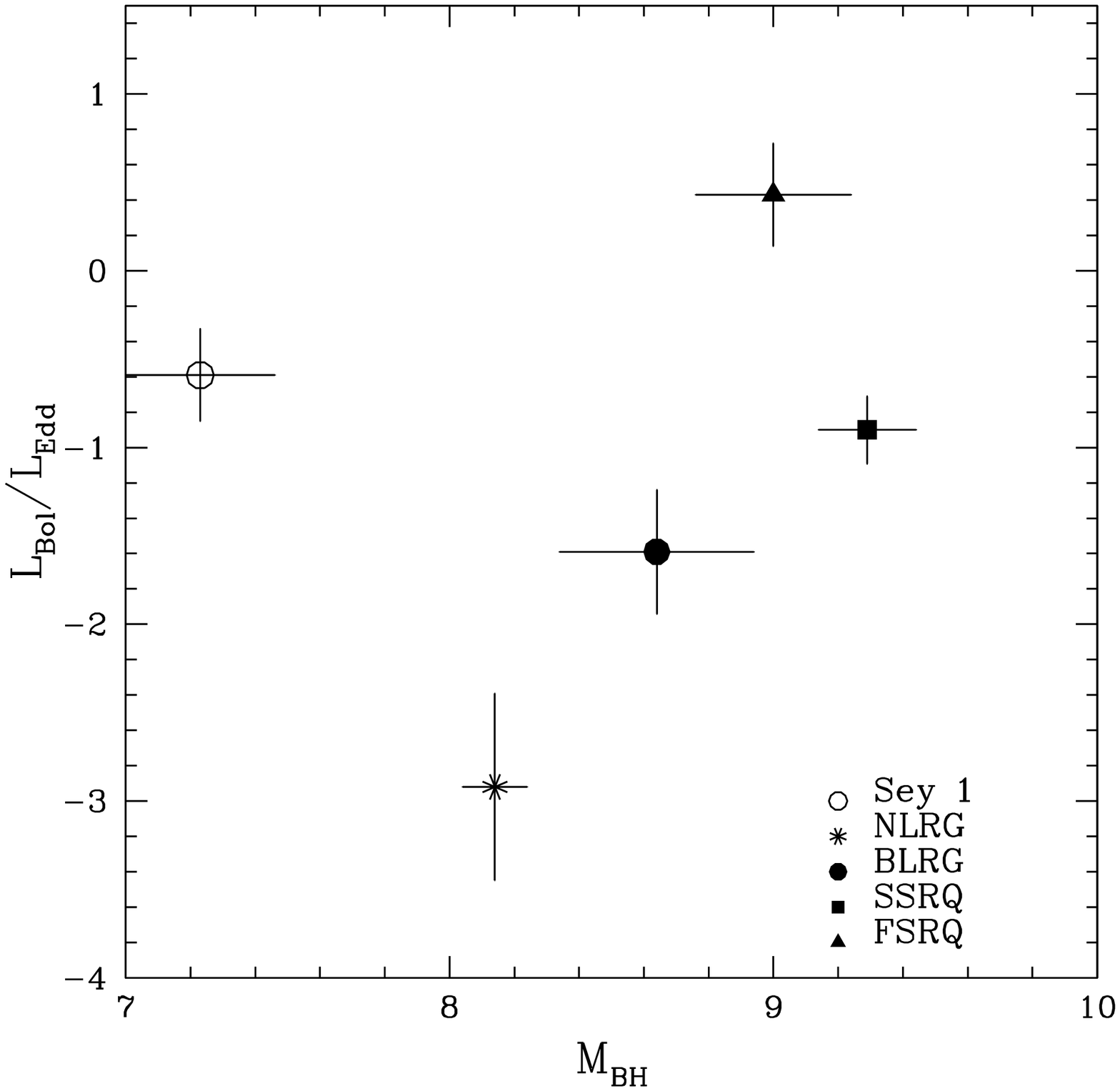} 
\caption{({\it Left Panel}) --Clear positive trend between  $L_{Bol}/L_{Edd}$ and Radio Core Dominance, impling a 
overestimation of the accretion rate in objects with strong jet emission. 
({\it Right panel}) -- Broad Line Radio Galaxies  are characterized by a smaller $L_{Bol}/L_{Edd}$ than Seyfert 1s
in spite of possible jet contamination. The accretion in NLRG could be understimate bacause of
its intrinsic absorbption}. 
\label{EffvsR}
\end{figure}

As recently reported by \citet{pag04} and successively confirmed by \citet{zho05}, the Baldwin effect is also present in radio quiet AGN. Indeed, the  $EW-L_{\rm 2-10\;keV}$ correlation is also observed in the small Perola sample ($\alpha_{\rm K}=0.04$). The above mentioned authors suggest that the Fe line (or at least its narrow component) is produced by the molecular torus at parsec scales and that the observed Baldwin effect is caused by a decrease of the obscuring material covering factor, $C_{\rm f}$, (i.e. an  increase of the opening angle) with luminosity.
\citet{zho05} found a clear correlation between iron line EW and accretion rate, which is stronger that the $EW-L_{\rm 2-10\;keV}$ relation. Assuming the torus model proposed by \citet{kro88}, they could relate the torus covering factor to the accretion rate ($C_{\rm f}\propto\dot{m}^{-0.5}$), and in turn the accretion, to the EW ($EW \propto C_{\rm f}\propto \dot{m}^{-0.5}$).
Although it would be reasonable to extend this result to our sample, it is difficult to ascribe the difference between BLRG and Seyfert 1s shown in Figure \ref{EffvsR} to a larger opening angle of the gas/molecular torus in radio loud galaxies. 
The accretion rate in BLRG is smaller than in Seyferts (see Figure \ref{EffvsR}) and therefore a larger $C_{\rm f}$ 
should be measured. Actually, 5 BLRG out of 10 need an extra $N_{\rm H}$ in addition to the Galactic one, while no Seyfert 1 requires intrinsic absorber, suggesting a gas rich environment, distributed at all angles.
As also pointed out by \citet{sam99}, who analyzed a large sample of \rl AGN with \asca, large cold gas absorbing columns are rather common in BLRG. On the other hand, in the picture outlined above, a larger  $C_{\rm f}$ implies larger iron line EW values, which are, instead, not observed. 

\section{Conclusions}

\sax analysis of 35 \rl AGN has shown the existence of a variety of X-ray spectra. In general, the shape of the spectrum appears smooth in FSRQ and rich of features in BLRG. In SSRQ the analysis is limited by poor statistics. However in the only case with high signal to noise spectrum, some similarity with BLRG is evident.

FSRQ in our sample are characterized by a harder power law and by the general absence of reprocessed features.
This is in clear agreement with the well consolidated idea that the jet component dominates this class of AGN.

In the other \rl AGN, the traces of a non-thermal Doppler component are more elusive and the results controversial and not conclusive yet.
Neither the spectral slopes nor the high energy cutoffs seem to be related to Core Radio Dominance ($R$) parameter, indicator 
of the orientation of the beamed radiation with respect to the light of sight. However a possible anti-correlation between the reprocessed features (EW, reflection) and $R$ is found, although not statistically robust. 

A $EW-L_{\rm 2-10 }$ anti-correlation (Baldwin effect) is observed in \rl AGN. This could be simply interpreted in terms of jet dilution, if the case of 3C 373, in which beamed non-thermal and thermal radiation are untangled, is extrapolated to other sources. However, since Seyfert 1 galaxies strengthen the case for the anticorrelation, other causes apart the jet must exist which could weaken the Fe line EW in non-blazar AGN. 

As recently proposed by \citet{min04}, the Baldwin effect could be explained for example within a light bending model. 
If this model is extended to \rl AGN, the weaker iron lines and larger X-ray fluxes observed in BLRG could then be 
simply explained in terms of larger heights of the primary source above a cold untruncated disk.
Incidentally, we note that a reduction of the reprocessed features could be also explained by 
a mildly relativistic motion of the X-ray source (for example a hot corona) directed away from
the disk \citep{bel9a,bel9b,rey97}. 

Alternative explanations seem to be less supported by our results.
The significantly larger $L_{\rm Bol}/L_{\rm Edd}$ ratio observed in our reference sample of Seyfert 1 when compared to BLRG, are in agreement with the presence of a different accretion flow configuration at least in the inner regions of radio galaxies.
However, this hypothesis is disfavored by \xmm and \chandra observations showing that gravitationally redshifted iron lines, supposed to be produced very close to the black hole, are not so easily detectable in Seyfert 1 as previously thought.

The Seyfert-BLRG different accretion rates also question the possibility that a strong ionization of the upper layers of the geometrically thin disk inhibits the Fe line production in BLRG. In a reasonable Seyfert-like configuration with a flaring corona above a thin disk, the EW is in fact expected to decrease with the accretion rate, in contrast with our results.

BLRG have smaller accretion rates but smaller EW than Seyfert galaxies. This contradicts also a model recently proposed \citep{zho05}, which shows that higher accretion rates imply flatter torii (and, as a consequence, smaller EW).
In spite of this, the role of the circumnuclear matter (as the main reprocessing medium in addition/alternative to the accretion disk) deserves a deeper investigation. Physical/geometrical differences as well as metal abundances of the gas environment should be investigated in \rqu and \rl AGN.
This point could be better tested by comparing NLRG and Seyfert 2 galaxies, which are both strongly affected by the gas environment.

\acknowledgments

This research has made use of the NASA/IPAC Extragalactic Database (NED) which is operated by the Jet Propulsion Laboratory, 
California Institute of Technology, 
under contract with the National Aeronautics and Space Administration.
We thank P. Giommi for valuable discussions and E. Palazzi for help on a statistical problem.

\clearpage

%
%

\begin{deluxetable}{lcccccc}
\tabletypesize{\scriptsize}
\tablewidth{0pt}
\tablecaption{BeppoSAX \rl AGN\label{tab:sam}}
\tablehead
{
\colhead{Name} & \colhead{RA$_{\rm 2000}$\tablenotemark{(a)}}       & \colhead{Dec$_{\rm 2000}$\tablenotemark{(a)}}     & \colhead{$z$\tablenotemark{(a)}}
               & \colhead{N$_{\rm H}^{\rm Gal}$\tablenotemark{(b)}} & \colhead{$R$\tablenotemark{c}} & \colhead{class\tablenotemark{d}}
\\
\colhead{}     & \colhead{(hh mm ss.s)}          & \colhead{(hh mm ss)}           & \colhead{}
               & \colhead{($10^{20}$ cm$^{-2}$)}& \colhead{} & \colhead{}
}
\scriptsize
\startdata
3C 18            &00 40 50.5     &$+$10 03 23    &0.188  &5.45  &$0.053\tablenotemark{e}$          &BLRG / FR II\\
3C 57            &02 01 57.1     &$-$11 32 34    &0.669  &1.87  & \nodata                          &SSRQ     \\
0208$-$512       &02 10 46.2     &$-$51 01 02    &0.999  &3.17  & \nodata                          &FSRQ     \\
NRAO 140         &03 36 30.1     &$+$32 18 29    &1.258  &14.8  &$0.22\pm0.07\tablenotemark{f}$    &FSRQ    \\
OF$-$109         &04 07 48.4     &$-$12 11 37    &0.573  &3.81  &$0.53\pm0.16\tablenotemark{f}$    &FSRQ    \\
3C 111           &04 18 21.3     &$+$38 01 36    &0.049  &31.5  &$0.04\pm0.02\tablenotemark{f}$    &BLRG / FR II\\
3C 120           &04 33 11.1     &$+$05 21 16    &0.033  &11.1  &$0.67\tablenotemark{e}$           &BLRG / FR I \\
Pictor A         &05 19 49.7     &$-$45 46 44    &0.035  &4.16  &$0.03\pm0.01\tablenotemark{f}$    &BLRG / FR II\\
0528+134         &05 30 56.4     &$+$13 31 55    &2.060  &26.5  &$0.49\pm0.24\tablenotemark{f}$    &FSRQ     \\
3C 171           &06 55 14.8     &$+$54 09 00    &0.238  &6.12  &$0.0016\pm0.0005\tablenotemark{f}$&NLRG / FR II \\
3C 184.1         &07 43 01.3     &$+$80 26 26    &0.118  &3.22  & \nodata                          &NLRG / FR II \\
4C 71.07         &08 41 24.3     &$+$70 53 42    &2.172  &2.92  &$0.46\pm0.14\tablenotemark{f}$    &FSRQ      \\
OM $-$161        &11 39 10.7     &$-$13 50 44    &0.558  &3.59  &$0.42\pm0.13\tablenotemark{f}$    &SSRQ      \\
3C 273           &12 29 06.7     &$+$02 03 09    &0.158  &1.79  &$4.13\pm1.24\tablenotemark{f}$    &FSRQ    \\
3C 279           &12 56 11.1     &$-$05 47 22    &0.536  &2.21  &$2.69\tablenotemark{e}$           &FSRQ    \\
PKS 1355$-$41    &13 59 00.2     &$-$41 52 53    &0.313  &5.61  &$0.03\tablenotemark{e}$           &SSRQ    \\
3C 300           &14 23 01.0     &$+$19 35 17    &0.270  &2.37  &$0.0055\pm0.0017\tablenotemark{f}$&NLRG / FR II\\
PKS 1510$-$08    &15 12 50.5     &$-$09 06 00    &0.360  &7.96  &$0.91\tablenotemark{e}$           &FSRQ  \\
PG 1512+370      &15 14 43.0     &$+$36 50 50    &0.371  &1.36  & \nodata                          &SSRQ  \\
RGB J1629+401    &16 29 01.3     &$+$40 08 00    &0.272  &0.90  & \nodata                          &FSRQ  \\
3C 345           &16 42 58.8     &$+$39 48 37    &0.593  &1.13  & \nodata                          &FSRQ  \\
RGB 1722+246     &17 22 41.2     &$+$24 36 19    &0.175  &4.95  & \nodata                          &BLRG / FR ? \\
4C 62.29         &17 46 14.0     &$+$62 26 55    &3.889  &3.36  &$3.33\pm1.01\tablenotemark{f}$    &FSRQ\\
3C 382           &18 35 02.1     &$+$32 41 50    &0.058  &7.75  &$0.05\pm0.02\tablenotemark{f}$     &BLRG / FR II    \\
3C 390.3         &18 42 09.0     &$+$79 46 17    &0.056  &4.24  &$0.09\pm0.03\tablenotemark{f}$       &BLRG / FR II    \\
4C +74.26        &20 42 37.3     &$+$75 08 02    &0.104  &11.9  &$0.95\pm0.25\tablenotemark{f}$       &SSRQ    \\
S5 2116+81       &21 14 01.2     &$+$82 04 48    &0.084  &7.39  &$1.78\pm0.56\tablenotemark{f}$       &BLRG / FR ?     \\
OX $-$158        &21 37 45.2     &$-$14 32 56    &0.200  &4.73  &$0.10\tablenotemark{e}$              &SSRQ     \\
PKS 2149$-$306   &21 51 55.5     &$-$30 27 54    &2.345  &2.12  & \nodata                             &FSRQ   \\
PKS 2153$-$69    &21 57 06.0     &$-$69 41 24    &0.028  &2.50  &$0.03\tablenotemark{e}$              &BLRG / FR ?  \\
3C 445           &22 23 49.6     &$-$02 06 12    &0.056  &5.01  &$0.04\tablenotemark{e}$              &BLRG / FR II    \\
3C 446           &22 25 47.2     &$-$04 57 01    &1.404  &5.50  & \nodata                             &FSRQ\\
PKS 2230+11      &22 32 36.4     &$+$11 43 51    &1.037  &5.04  &$2.11\pm0.63\tablenotemark{f}$       &FSRQ    \\
PKS 2243$-$123   &22 46 18.2     &$-$12 06 51    &0.632  &4.65  & \nodata                             &FSRQ     \\
3C 454.3         &22 53 57.7     &$+$16 08 54    &0.859  &6.50  &$15\pm5.00\tablenotemark{f}$         &FSRQ    \\
\enddata
\tablenotetext{a}{NASA Extragalactic Database, NED}
\tablenotetext{b}{Dickey \& Lockman 1990}
\tablenotetext{c}{Radio Core Dominance $R=S_{\rm core}/(S_{\rm tot}-S_{\rm core})$ at 5 GHz}
\tablenotetext{d}{Radio/optical classification}
\tablenotetext{e}{Morganti et al. (1993)}
\tablenotetext{f}{Fan \& Zhang (2003)}
\end{deluxetable}

\begin{deluxetable}{ccccc}
\tabletypesize{\scriptsize}
\tablewidth{0pt}
\tablecaption{BeppoSAX Journal \label{tab:jou}}
\tablehead
{
\colhead{Source Name}       & \colhead{Obs. Date}     & \colhead{T$_{\rm MECS}$\tablenotemark{(a)}} 
                            & \colhead{T$_{\rm LECS}$\tablenotemark{(a)}}        & \colhead{T$_{\rm PDS}$\tablenotemark{(b)}} 
\\
\colhead{}                  & \colhead{(yyyy-mm-dd)}  & \colhead{(s)}            & \colhead{(s)}            & \colhead{(s)} 
}
\startdata
3C 18              &1998-12-17    &47629       &20183  &N.S.\\
3C 57              &1999-01-27    &18388       &5164   &N.S.\\
                   &1999-01-28    &21939       &9815   &N.S.\\
0208-512           &2001-01-14    &34256       &15893  &N.S.\\
NRAO 140           &1999-08-05    &49987       &17528  &22924\\
OF $-$109          &1996-09-22    &16755       &0      &N.S.\\
3C 111             &1998-03-08    &69612       &21048  &32972\\
3C 120             &1997-09-20    &81596       &34695  &35751\\
Pictor A           &1996-12-10    &14746       &5272   &N.S.\\
                   &2001-01-15    &37841       &28324  &N.S.\\
0528+134           &1997-02-21    &14435       &6321   &N.S.\\
                   &1997-02-22    &13304       &5045   &N.S.\\
                   &1997-02-27    &7318        &4423   &N.S.\\
                   &1997-03-01    &13522       &4261   &N.S.\\
                   &1997-03-03    &14080       &5052   &N.S.\\
                   &1997-03-04    &11240       &2793   &N.S. \\
                   &1997-03-06    &12659       &3082   &N.S. \\
                   &1997-03-11    &11459       &2496   &N.S.\\
3C 171             &2001-04-27    &101656      &N.S.   &N.S\\
3C 184.1           &2001-01-27    &54069       &N.S.   &N.S.     \\
4C 71.07           &1998-05-27    &42639       &18518  &16493\\
OM $-$161          &1997-01-11    &28009       &13783  &N.S.\\
3C 273             &1996-07-18    &129887      &11849  &60636\\
                   &1997-01-13    &25136       &13603  &11436\\
                   &1997-01-15    &24014       &13286   &10848\\
                   &1997-01-17    &27349       &12474  &12471\\
                   &1997-01-22    &22289       &8779   &9277\\
                   &1998-06-24    &72151       &27941  &33820\\
                   &2000-01-09    &85165       &34782  &42213\\
                   &2000-06-13    &68082       &29586  &31940\\
                   &2001-06-12    &38440       &16909  &18495\\
3C 279             &1997-01-13    &21800       &7429   &9034\\
                   &1997-01-15    &23906       &8398   &9991\\
                   &1997-01-18    &2581        &523    &N.S.\\
                   &1997-01-21    &11747       &4188   &5083\\
                   &1997-01-23    &24698       &11161  &11423\\
PKS $1355-41$      &1998-01-10    &21925       &4548   &N.S.\\
3C 300             &2002-01-12    &157267      &56756  &N.S. \\
PKS $1510-089$     &1998-03-08    &43869       &16155  &N.S.\\
PG 1512+37         &1998-01-25    &18708       &5125   &N.S.\\
RGB J1629+4008     &1999-08-11    &44899       &21221  &N.S.    \\
3C 345             &1999-02-19    &25866       &11107  &N.S.\\
RGB J1722+246      &2000-02-13    &44064       &12360  &N.S.\\
4C 62.29           &1997-03-29    &143032      &58449  &N.S.\\
3C 382             &1998-09-20    &85645       &35181  &41613\\
3C 390.3           &1997-01-09    &100618      &34832  &45720\\
4C +74.26          &1999-05-17    &100276      &44499  &47410\\
S5 2116+81         &1998-04-29    &28946       &13390  &13296\\
                   &1998-10-12    &19674       &5666   &10488\\
OX $-158$          &1996-10-29    &14235       &4704   &N.S.\\
PKS $2149-306$     &1997-10-31    &39439       &17862  &16677\\
PKS $2152-69$      &1996-09-29    &16872       &0      &N.S.\\
3C 445             &1999-11-30    &99883       &30062  &46472\\
3C 446             &1997-11-10    &16175       &9362   &N.S.\\
PKS 2230+114       &1997-11-11    &23811       &11008  &11597\\
                   &1997-11-13    &22256       &10286  &N.S.\\
                   &1997-11-16    &25748       &13244  &12118\\
                   &1997-11-18    &12237       &7030   &5354\\
                   &1997-11-21    &19376       &9489   &8612\\
PKS 2243-123       &1998-11-18    &27492       &10236  &N.S.\\
3C 454.3           &2000-06-05    &48514       &17693  &22439\\
\enddata
\tablenotetext{(a)}{On source time}
\tablenotetext{(b)}{N.S. indicates no useful spectrum}
\end{deluxetable}

\begin{deluxetable}{lcc ccc ccc}
\tabletypesize{\scriptsize}
\tablewidth{0pt}
\tablecaption{Spectral fit results \label{tab:fit}}
\tablehead
{
\colhead{Source Name}       & \colhead{$\Gamma$}                               & \colhead{E$_{\rm C}$} & \colhead{F$_{\rm 2-10\;keV}$\tablenotemark{(a)}} 
                            & \colhead{Log (L$_{\rm 2-10\;keV})$\tablenotemark{(a)}} & \colhead{N$_{\rm H}$} & \colhead{Refl.}
                            & \colhead{EW}                                     & \colhead{$\chi^2$/dof}
\\
                            &                                                  & \colhead{(keV)}                       & \colhead{($10^{-11}$ erg cm$^{-2}$ s$^{-1}$)} 
                            & \colhead{(erg s$^{-1}$)}                     & \colhead{($10^{22}$ cm$^{-2}$)} & \colhead{}
                            & \colhead{(eV)}                                   & \colhead{}
}
\tablecolumns{9}
\startdata
\cutinhead{BLRG} 
3C 18          & 
$1.86^{+0.17}_{-0.15}$   & \nodata            & $0.26^{+0.08}_{-0.05}$ & 44.28 & $0.36^{+0.23}_{-0.16}$        & \nodata             & $<314$               & 60/52   \\
3C 111\tablenotemark{(b)} & 
$1.58^{+0.04}_{-0.06}$   & $146^{+224}_{-68}$ & $2.71^{+0.22}_{-0.18}$ & 44.08 & $0.55^{+0.09}_{-0.08}$        & $<0.3$              & $<72$                & 124/114 \\
3C 120\tablenotemark{(c)}& 
$1.78^{+0.09}_{-0.13}$   & $105^{+106}_{-41}$ & $4.9^{+0.5}_{-0.9}$    & 44.00 & $0.05^{+0.04}_{-0.03}$        & $0.5^{+0.5}_{-0.3}$ & $46^{+35}_{-38}$     & 132/128 \\
Pictor A       & 
$1.64^{+0.05}_{-0.06}$   & \nodata            & $1.10^{+0.10}_{-0.10}$ & 43.48 & \nodata                       & \nodata             & $<148$               & 97/108  \\
RGB J1722+246   & 
$1.82^{+0.28}_{-0.29}$   & \nodata            & $0.10^{+0.03}_{-0.03}$ & 43.78 & \nodata                       & \nodata             & $<807$               & 19/19   \\
3C 382\tablenotemark{(c)}& 
$1.77^{+0.05}_{-0.08}$   & $138^{+92}_{-41}$  & $5.98^{+0.20}_{-0.39}$ & 43.48 & \nodata                       & $0.3^{+0.3}_{-0.1}$ & $39^{+26}_{-23}$     & 143/144\\
3C 390.3       & 
$1.72^{+0.04}_{-0.04}$   & $257^{+987}_{-121}$& $2.20^{+0.06}_{-0.05}$ & 44.11 & $0.07^{+0.02}_{-0.02}$        & $0.7^{+0.4}_{-0.3}$ & $121^{+29}_{-35}$    & 124/117 \\
S5 2116+81     & 
$1.80^{+0.08}_{-0.06}$   & $>112$             & $1.50^{+0.09}_{-0.13}$ & 44.30 & $0.05^{+0.04}_{-0.02}$        & $<1.0$              & $<95$                & 105/109 \\
PKS 2152$-$69    & 
$1.83^{+0.12}_{-0.12}$   & \nodata            & $0.72^{+0.10}_{-0.10}$ & 43.00 & \nodata                       & \nodata             & $<336$               & 55/37   \\
3C 445         & 
$1.73^{+0.47}_{-0.50}$   & $>44$              & $1.5^{+1.5}_{-0.8}$    & 43.95 & $32^{+58}_{-17}$\tablenotemark{(d)}     & $1.2^{+2.5}_{-1.1}$& $170^{+219}_{-95}$& 66/63 \\
 &                       &                    &                        &       & $6.8^{+5.6}_{-3.6}$\tablenotemark{(e)}  & \nodata  & \nodata              &       \\
\cutinhead{NLRG}
3C 171         & 
$2.2^{+0.9}_{-0.5}$      & \nodata            & $0.07^{+2.59}_{-0.04}$ & 43.97 & 9.9$^{+16.8}_{-7.6}$          & \nodata             & $<789$               & 4/6     \\
3C 184.1       & 
$1.65^{+0.39}_{-0.36}$   & \nodata            & $0.45^{+0.57}_{-0.24}$ & 44.09 & 10.6$^{+3.8}_{-3.1}$          & \nodata             & $<292$               & 34/33   \\
3C 300         & 
1.37$^{+0.52}_{-0.48}$   & \nodata            & $1.39^{+0.52}_{-0.50}$ & 43.45 & $<1.14$                       & \nodata             & $<649$               & 13/17   \\
\cutinhead{SSRQ}
3C 57          & 
$1.85^{+0.14}_{-0.14}$   & \nodata            & $0.23^{+0.05}_{-0.06}$ & 45.36 & \nodata                       & \nodata             & $<148$               & 155/137 \\
OM $-$161$^a$  & 
$1.74^{+0.14}_{-0.13}$   & \nodata            & $0.26^{+0.07}_{-0.05}$ & 45.24 & \nodata                       & \nodata             & $<230$               & 43/50   \\
PKS 1355$-$41  & 
$1.81^{+0.09}_{-0.10}$   & \nodata            & $0.48^{+0.06}_{-0.07}$ & 44.99 & \nodata                       & \nodata             & $<178$               & 47/48   \\
OX $-$158      & 
$1.78^{+0.09}_{-0.07}$   & \nodata            & $0.86^{+0.08}_{-0.09}$ & 44.84 & \nodata                       & \nodata             & $<387$               & 64/50   \\
4C +74.26\tablenotemark{(f)}  & 
$1.83^{+0.09}_{-0.07}$   & $145^{+294}_{-66}$ & $2.40^{+0.37}_{-0.15}$ & 44.68 & $0.25^{+0.06}_{-0.05}$        & $1.3^{+0.9}_{-0.7}$ & $45^{+44}_{-45}$     & 120/105 \\
PG 1512+37     & 
$1.84^{+0.17}_{-0.24}$   & \nodata            & $0.19^{+0.10}_{-0.07}$ & 44.76 & \nodata                       & \nodata             & $<362$               & 19/20   \\
\cutinhead{FSRQ}
0208$-$512\tablenotemark{(g)}& 
$1.62^{+0.12}_{-0.12}$   & \nodata            & $0.48^{+0.13}_{-0.10}$ & 45.96 & \nodata                       & \nodata             & $<62$                & 39/43   \\
NRAO 140\tablenotemark{(b)}& 
$1.63^{+0.15}_{-0.29}$   &       $>34$        & $0.73^{+0.14}_{-0.14}$ & 46.38 & $0.18^{+0.14}_{-0.12}$        &  $<0.3$             & $<64$                & 71/64   \\
OF $-$109      & 
$1.78^{+0.15}_{-0.15}$   & \nodata            & $0.39^{+0.11}_{-0.10}$ & 45.43 &\nodata                        & \nodata             & $<263$               & 42/39   \\
0528+134       & 
$1.36^{+0.08}_{-0.09}$   & \nodata            & $0.25^{+0.05}_{-0.05}$ & 46.20 & \nodata                       & \nodata             & $<35$                & 47/50   \\
4C 71.07\tablenotemark{(h)}& 
$1.34^{+0.03}_{-0.02}$   &    $>346$          & $2.65^{+0.35}_{-0.23}$ & 47.25 & $1.1^{+1.6}_{-0.8}$           &  $<0.2$             & $<34$                & 106/106 \\
3C 273\tablenotemark{(c, i)}& 
$1.63^{+0.05}_{-0.03}$   & $957^{+752}_{-136}$& $8.88^{+0.10}_{-0.08}$ & 45.63 & 0.09$^{+0.003}_{-0.005}$     &0.14$^{+0.05}_{-0.03}$& $22^{+34}_{-11}$     & 144/127 \\
3C 279         & 
$1.68^{+0.09}_{-0.05}$   & $>95$              & $0.58^{+0.04}_{-0.04}$ & 45.53 & \nodata                       &  $<0.5$             & $<29$                & 228/226 \\
PKS 1510$-$089 & 
$1.36^{+0.09}_{-0.10}$   & \nodata            & $0.53^{+0.09}_{-0.09}$ & 45.08 & \nodata                       & \nodata             & $<184$               & 34/48   \\
RGB J1629+4008 & 
$2.58^{+0.11}_{-0.12}$   & \nodata            & $0.12^{+0.03}_{-0.02}$ & 44.34 & \nodata                       & \nodata             & $<413$               & 25/27   \\
3C 345         & 
$1.59^{+0.10}_{-0.11}$   & \nodata            & $0.51^{+0.10}_{-0.10}$ & 45.53 & \nodata                       & \nodata             & $<127$               & 62/50   \\
4C 62.29       & 
$1.62^{+0.17}_{-0.16}$   & \nodata            & $0.06^{+0.04}_{-0.02}$ & 46.30 & \nodata                       & \nodata             & $<300$               & 31/46   \\
PKS 2149$-$306 & 
$1.35^{+0.14}_{-0.12}$   &  $>141$            & $0.80^{+0.08}_{-0.09}$ & 46.82 & \nodata                       & $<0.2$              & $<92$                & 69/80   \\
3C 446         & 
$1.74^{+0.36}_{-0.35}$   & \nodata            & $0.12^{+0.24}_{-0.07}$ & 45.59 & \nodata                       & \nodata             & $<392$               & 26/19   \\
PKS 2230+114   & 
$1.52^{+0.09}_{-0.08}$   & $>103$             & $0.61^{+0.05}_{-0.07}$ & 46.09 & \nodata                       & $<0.3$              & $<74$                & 141/129 \\
PKS 2243$-$123 & 
$1.70^{+0.12}_{-0.23}$   & \nodata            & $0.18^{+0.09}_{-0.06}$ & 45.18 & \nodata                       & \nodata             & $<172$               & 23/25   \\
3C 454.3       & 
$1.37^{+0.15}_{-0.11}$   & $>172$             & $1.13^{+0.20}_{-0.13}$ & 46.15 & $0.5^{+0.5}_{-0.4}$           & $<0.5$              & $<40$                & 79/73   \\
\enddata
\tablenotetext{(a)}{Unabsorbed Flux/Luminosity}
\tablenotetext{(b)}{The absorber is not intrisic to the source, \nh is obtained assuming $z=0$}
\tablenotetext{(c)}{Source with soft component}
\tablenotetext{(d)}{Covering factor C$_{\rm f}=0.67^{+1.12}_{-0.19}$}
\tablenotetext{(e)}{Covering factor C$_{\rm f}=0.88^{+0.05}_{-0.09}$}
\tablenotetext{(f)}{The fit improves if all Fe line parameters are left free to vary ($\chi^2/dof=113/103$)}
\tablenotetext{(g)}{An instrumental line at 5.1 keV is included in the fit (Tavecchio et al. 2002)}
\tablenotetext{(h)}{N$_{\rm H}$ is associated to a foreground damped Ly$_{\alpha}$ absorber located at $z=0.3130$ (Turnshek et al. 2003)}
\tablenotetext{(i)}{The uncertainties are calculated fixing the PDS/MECS inter-calibration factor to the best fit value}
\end{deluxetable}

\begin{deluxetable}{ll ccc}
\tabletypesize{\scriptsize}
\tablewidth{0pt}
\tablecaption{Soft component \label{tab:soft}}
\tablehead
{
\colhead{Model}  & \colhead{Source Name}    &  \colhead{kT}   & \colhead{F$_{\rm 0.2-2\;keV}$\tablenotemark{(a)}} & \colhead{Log(L$_{\rm 0.2-2\;keV}$)\tablenotemark{(a)}} 
\\
\colhead{}       & \colhead{}               &  \colhead{(keV)}& \colhead{(10$^{-12}$ erg cm$^{-2}$ s$^{-1}$)}        & \colhead{(10$^{43}$ erg s$^{-1}$)} 
}
\tablecolumns{5}
\startdata
Raymond Smith    & 3C 120                   & $1.2^{+0.4}_{-0.3}$                & 7.5$^{+5.8}_{-3.5}$     & 43.20     \\
Raymond Smith    & 3C 382\tablenotemark{(b)}& $1.2^{+0.2}_{-0.4}$                & 6.4$^{+4.9}_{-3.7}$     & 43.61     \\
                 &                          & 0.16$^{+0.03}_{-0.04}$             & 12.9$^{+7.9}_{-4.4}$    & 43.92     \\
Black Body       & 3C 273                   & (6.1$^{+0.6}_{-0.9})\times10^{-2}$ & 9.9$^{+0.7}_{-4.3}$     & 44.80     \\
\enddata
\tablenotetext{(a)}{Unabsorbed Flux/Luminosity}
\tablenotetext{(b)}{3C 382 requires two soft components (Grandi et al. 2001)}
\end{deluxetable}

\clearpage
\begin{landscape}
\begin{deluxetable}{l ccc ccc ccc ccc cc}
\tabletypesize{\tiny}
\tablewidth{0pt}
\tablecaption{Average Properties \label{tab:avg}}
\tablehead
{
\colhead{Class} & \mcl{2}{c}{Spectral index}  
                & \mcl{2}{c}{Energy cut-off (keV)} 
                & \mcl{2}{c}{Reflection} 
                & \mcl{2}{c}{Eq. Width (eV)}
                & \mcl{2}{c}{L$_{\rm 0.2-2\;keV}$}
                & \mcl{2}{c}{Redshift $z$}
                & \mcl{2}{c}{Beaming $R$} \\
\colhead{}      & \colhead{$<\Gamma>$}         &     \colhead{$\sigma$\tnm{a}} 
                & \colhead{$<E_{\rm cut}>$}    &     \colhead{$\sigma$\tnm{a}} 
                & \colhead{$<Refl.>$}          &     \colhead{$\sigma$\tnm{a}}
                & \colhead{$<EW>$}             &     \colhead{$\sigma$\tnm{a}} 
                & \colhead{$<Log({\rm L})>$} &     \colhead{$\sigma$\tnm{a}}   
                & \colhead{$<z>$}              &     \colhead{$\sigma$\tnm{a}} 
                & \colhead{$<R>$}              &     \colhead{$\sigma$\tnm{a}} 
}
\tablecolumns{15}
\startdata
BLRG        & $1.75\pm0.03$ & 0.09 & $163\pm33$                 & 66  & $0.68\pm0.19$                 & 0.39 & $94\pm32$               & 63  & $43.96\pm0.14$ & 0.45 & $0.08\pm0.02$   & 0.06 & $-1.04\pm0.21$ & 0.63 \\
BLRG\tnm{b} & \nd           & \nd  & \nd                        & \nd & $0.50\pm0.12$                 & 0.21 & $69\pm26$               & 46  & \nd            & \nd  & \nd             & \nd  & \nd            & \nd  \\
NLRG        & $1.74\pm0.24$ & 0.42 & \nd                        & \nd & \nodata                       & \nd  & \nd                     & \nd & $43.84\pm0.20$ & 0.34 & $0.21\pm0.05$   & 0.09 & $-2.53\pm0.27$ & 0.38 \\
SSRQ\tnm{c}        & $1.81\pm0.02$ & 0.04 & $145^{+294}_{-66}$  & \nd & $1.3^{+0.9}_{-0.7}$    & \nd  & $45^{+44}_{-45}$ & \nd & $44.98\pm0.11$ & 0.27 & $0.37\pm0.09$   & 0.22 & $-0.73\pm0.33$ & 0.66 \\
FSRQ\tnm{d}        & $1.55\pm0.04$ & 0.16 & $957^{+752}_{-136}$ & \nd & $0.14^{+0.05}_{-0.03}$ & \nd  & $22^{+34}_{-11}$ & \nd & $45.94\pm0.16$ & 0.60 & $1.26\pm0.26$   & 1.01 & \phantom{\phs}$0.16\pm0.18$ & 0.57\\
Sey         & $1.79\pm0.05$ & 0.14 & $166\pm30$                 & 73  & $0.75\pm0.11$                 & 0.30 & $137\pm18$              & 48  & $43.44\pm0.18$ & 0.45 & $0.018\pm0.003$ & 0.08 & \nd            & \nd \\
\enddata
\tablenotetext{(a)}{1$\sigma$ distribution spread}
\tablenotetext{(b)}{Excluding 3C 445}
\tablenotetext{(c)}{Energy cut-off, Reflection and Equivalent Width refer to 4C+74.26 only}
\tablenotetext{(d)}{Energy cut-off, Reflection and Equivalent Width refer to 3C 273 only}
\end{deluxetable}
\clearpage
\end{landscape}

\begin{deluxetable}{lcccc}
\tabletypesize{\scriptsize}
\tablewidth{0pt}
\tablecaption{Black hole masses and accretion rates \label{tab:bhm}}
\tablehead
{
\colhead{Source Name}  & \colhead{Log ($M_{\rm BH}$)}    &  \colhead{Log($L_{\rm Bol}$)} & \colhead{Log (L$_{\rm Bol}/L_{\rm Edd}$)}  & \colhead{Ref.\tablenotemark{(a)}}
\\
\colhead{}             & \colhead{M$_{\odot}$}            & \colhead{(erg s$^{-1}$)}         
& \colhead{} & \colhead{}
}
\tablecolumns{5}
\startdata
\cutinhead{\it BLRG}
3C 18            &8.92   & 44.45 &  $-2.57$  & M\\
3C 111           &9.56   & 45.58 &  $-2.08$  & M\\
3C 120           &7.42   & 45.34 &  $-0.18$  & W\\
3C 382           &9.06   & 45.55 &  $-1.61$  & M\\
3C 390.3         &8.55   & 44.88 &  $-1.77$  & W\\
3C 445           &8.33   & 45.10 &  $-1.33$  & M\\
\cutinhead{\it NLRG}
3C 171           &7.96   & 42.33 &  $-3.73$ & M \\
3C 184.1         &8.32   & 44.00 &  $-2.42$ & M \\
3C 300           &8.14   & 43.62 &  $-2.62$ & M \\
\cutinhead{\it SSRQ}
3C 57            &9.27   & 46.84 &  $-0.53$ & W \\
OM $-$161        &8.78   & 46.78 &  $-0.10$ & W\\
PKS 1355$-$41    &9.73   & 46.48 &  $-1.35$ & W\\
OX $-$158        &8.94   & 46.17 &  $-0.87$ & W\\
4C +74.26        &9.62   & 46.23 &  $-1.49$ & W\\
PG 1512+370      &9.41   & 46.46 &  $-1.05$ & B\\
\cutinhead{\it FSRQ}
0208$-$512       &9.21   & 48.60 &  \phantom{\phs}$1.19$ & F$^b$\\
OF$-$109         &9.47   & 47.40 &  $-0.17$ & W\\
0528+134         &9.40   & 49.65 &  \phantom{\phs}$2.15$ & X \\
4C 71.07         &9.45   & 48.41 &  \phantom{\phs}$0.77$ & F$^b$ \\
3C 273           &9.00   & 47.10 &  $-3.7\times10^{-4}$ & X    \\
3C 279           &8.43   & 46.10 &  $-0.43$ & W\\
PKS 1510$-$08    &8.65   & 46.38 &  $-0.37$ & W \\
3C 345           &9.42   & 46.89 &  $-0.63$ & W\\
3C 446           &7.90   & 47.10 &  \phantom{\phs}$1.10$ & X\\
PKS 2230+11      &9.00   & 48.16 &  \phantom{\phs}$1.06$ & X\\
3C 454.3         &9.17   & 47.27 &  $-3.7\times10^{-4}$ & W \\
\cutinhead{\it Seyfert 1}
NGC 3783         &6.94    & 44.41 & $-0.63$ & W\\
IC 4329A         &6.77    & 44.78 & $-0.09$ & W\\
NGC 5548         &8.03    & 44.83 & $-1.30$ & W\\
MKN 509          &7.86    & 45.03 & $-0.93$ & W\\
NGC 7469         &6.84    & 45.28 & $-0.34$ & W \\
NGC 4593         &6.91    & 44.09 & $-0.92$ & W\\
\cutinhead{\it Average Values\tablenotemark{(c) }}
BLRG            & 8.64$\pm$0.30 (0.73) &45.15$\pm$0.18 (0.44)& $-1.59\pm0.35$ (0.81) & \\
NLRG            & 8.14$\pm$0.10 (0.18) &43.32$\pm0.50$ (0.88)& $-2.92\pm0.53$ (0.71) & \\
SSRQ            & 9.29$\pm$0.15 (0.37) &46.49$\pm$0.11 (0.28)& $-0.90\pm0.19$ (0.52) & \\
FSRQ            & 9.01$\pm$0.15 (0.50) &47.55$\pm$0.32 (1.05)& \phantom{\phs}$0.44\pm0.27$ (0.89) & \\
Seyfert 1        & 7.23$\pm$0.23 (0.56) &44.74$\pm$0.18 (0.43)& $-0.59\pm0.25$ (0.61) & \\
\enddata     
\tablenotetext{(a)}{M: Marchesini, Celotti \& Ferrarese (2004) W: Woo\& Urry (2002); B: Bian \& Zhao (2004); F: Fan \& Cao (2004); X: Xie, Zhou \& Liang (2004)}
\tablenotetext{(b)}{Bolometric luminosity is assumed to be $\sim 2L_{\gamma}\sim2(L_{\gamma, min}+ L_{\gamma, max}/2)$, being $L_{\gamma, min}$ and $L_{\gamma, max}$, the minimum and maximum $\gamma$ luminosities
reported by Fan \& Cao (2004) in Table 1.}
\tablenotetext{(c)}{For each quantity, mean $\pm$ standard error on the mean and 
standard deviation (put in parenthesis) are reported.
}
\end{deluxetable}

\clearpage

%
%



\clearpage


\end{document}